\documentclass[prd,aps,preprint,showpacs,nofootinbib]{revtex4-1}
\usepackage{epsfig}
\usepackage{amsmath}
\usepackage{rotating}

\newcommand{\nslash}{\kern 0.2 em n\kern -0.50em /}
\newcommand{\kslash}{\kern 0.2 em k\kern -0.45em /}
\newcommand{\pslash}{\kern 0.2 em p\kern -0.50em /}
\newcommand{\Sslash}{\kern 0.2 em S\kern -0.50em /}
\newcommand{\Pslash}{\kern 0.2 em P\kern -0.50em /}
\newcommand{\Dslash}{\kern 0.2 em D\kern -0.65em /\kern 0.15em}

\newcommand{\de}{d}                    
\newcommand{\xbj}{x}                   

\newcommand{\slim}{\mskip 1.5mu}              

\newcommand{\bkt}{\bm{k}_{\scriptscriptstyle T}}

\newcommand{\bpt}{\bm{p}_{\scriptscriptstyle T}}
\newcommand{\bP}{\bm{P}}

\newcommand{\hF}{D}

\newcommand{\ba}{\begin{eqnarray}}
\newcommand{\ea}{\end{eqnarray}}
\newcommand{\be}{\begin{equation}}
\newcommand{\ee}{\end{equation}}

\begin{document}

\title{Transverse Momentum Dependent Parton
Distribution/Fragmentation Functions at an Electron-Ion
Collider}

\author{\centering  M.~Anselmino$^1$, H.~Avakian$^2$, D.~Boer$^3$, F.~Bradamante$^4$,
M.~Burkardt$^5$, J.P.~Chen$^2$, E.~Cisbani$^6$, M.~Contalbrigo$^7$, D.~Crabb$^8$, D.~Dutta$^9$, L.~Gamberg$^{10}$, H.~Gao$^{11}$, D.~Hasch$^{12}$,
J.~Huang$^{13}$, M.~Huang$^{11}$, Z.~Kang$^{14}$, C.~Keppel$^{15}$, G.~Laskaris$^{11}$, Z-T.~Liang$^{16}$, M.X.~Liu$^{17}$, N.~Makins$^{18}$, R.D.~Mckeown$^{19}$, A.~Metz$^{20}$, Z-E.~Meziani$^{20}$, B.~Musch$^2$, J-C.~Peng$^{18}$, A.~Prokudin$^2$,
X.~Qian$^{19}$, Y.~Qiang$^2$, J.W.~Qiu$^{21}$, P.~Rossi$^{12}$, P.~Schweitzer$^{22}$, J.~Soffer$^{20}$, V.~Sulkosky$^2$, Y.~Wang$^{23}$, B.~Xiao$^{24}$, Q.~Ye$^{11}$, Q-J.~Ye$^{11}$, F.~Yuan$^{24}$, X.~Zhan$^{25}$, Y.~Zhang$^2$, W.~Zheng$^{11}$, J.~Zhou$^{20}$,  \\
\centering 
$^1${\it Universita di Torino and INFN, Sezione di Torino, I-{10}125, Torino, Italy} \\
$^2${\it Thomas Jefferson National Accelerator Facility\\
12000 Jefferson Avenue, Newport News, Virginia 23606, USA}\\
$^3${\it KVI, University of Groningen, NL-9747 AA Groningen, The Netherlands}\\
$^4${\it Dipartimento di Fisica, Universit{\`a} degli Studi di Trieste, and INFN, Sezione di Trieste, 34127 Trieste, Italy}\\
$^5${\it New Mexico State University, Las Cruces, NM 88003, USA}\\
$^6${\it Istituto Nazionale di Fisica Nucleare, Sezione Roma 1, Gruppo Sanit\`a and Physics Laboratory, Istituto Superiore di Sanit\`a, 00161 Roma, Italy}\\
$^7${\it Istituto Nazionale di Fisica Nucleare, Sezione di Ferrara and Dipartimento di Fisica, Universit\`a di Ferrara, 44100 Ferrara, Italy}\\
$^8${\it University of Virginia, Charlottesville, Virginia 22901}\\
$^9${\it Mississippi State University, Starkeville, Mississippi 39762, USA}\\
$^{10}${\it Penn State-Berks, Reading, PA 19610, USA \\}
$^{11}${\it Triangle Universities Nuclear Laboratory and \\
Department of Physics, Duke University, Durham, NC 27708, USA\\}
$^{12}${\it Istituto Nazionale di Fisica Nucleare, Laboratori Nazionali di Frascati, 00044 Frascati, Italy}\\
$^{13}${\it Massachusetts Institute of Technology, Cambridge, Massachusetts 02139, USA}\\
$^{14}${\it RIKEN BNL Research Center, Brookhaven National Laboratory, Upton, NY 11973, USA}\\
$^{15}${\it Hampton University, Hampton, Virginia 23668, USA}\\
$^{16}${\it School of Physics, Shandong University, Jinan, Shandong 250100, China}\\
$^{17}${\it Physics Division, Los Alamos National Laboratory, Los Alamos, NM 87545}\\
$^{18}${\it University of Illinois, Urbana, IL 61801}\\
$^{19}${\it California Institute of Technology, Pasadena, CA 91125, USA\\}
$^{20}${\it Temple University, Philidalphia, PA 19122, USA \\}
$^{21}${\it Physics Department, Brookhaven National Laboratory, Upton, NY 11973, USA}\\
$^{22}${\it University of Connecticut, Storrs, CT 06269, USA}\\
$^{23}${\it Tsinghua University, Beijing 10084, China}\\
$^{24}${\it Lawrence Berkeley National Laboratory, Berkeley, CA, USA} \\
$^{25}${\it Physics Division, Argonne National Laboratory, Argonne, Illinois 60439}\\
}

\begin{abstract}
We present a summary of a recent workshop held at Duke University 
on Partonic Transverse Momentum in Hadrons: 
Quark Spin-Orbit Correlations and Quark-Gluon Interactions.
The transverse momentum dependent parton distribution functions (TMDs), parton-to-hadron fragmentation functions, and multi-parton correlation functions, were discussed extensively at the Duke workshop. In this paper,
we summarize first the theoretical issues concerning the study of partonic structure of hadrons at 
a future electron-ion collider (EIC) with emphasis on the TMDs.
We then present simulation results on
experimental studies of TMDs through measurements of single spin asymmetries (SSA) from semi-inclusive deep-inelastic scattering (SIDIS) processes  
with an EIC, and discuss the requirement of the detector for 
SIDIS measurements. 
The dynamics of parton correlations in the nucleon is further explored via a study of SSA in $D$ ($\bar{D}$) production at large transverse momenta with the aim of accessing the unexplored tri-gluon correlation functions.
The workshop participants identified the SSA measurements in SIDIS as a golden program to study TMDs in both the sea and valence quark regions and to study the role of gluons, with the 
Sivers asymmetry measurements as examples.
Such measurements will  
lead to major advancement in our understanding of TMDs in the valence quark region, and more importantly also allow for the
investigation of TMDs in the unexplored sea quark region along with a study of their evolution. 
\end{abstract}

 

\maketitle

\section{Introduction}

Understanding the internal structure of nucleon and nucleus in terms of quarks and gluons, the fundamental degrees of freedom of Quantum Chromodynamics (QCD) has been and still is the frontier of subatomic physics research.  QCD as a theory of the strong interaction has been well-tested by observables with a large momentum transfer in high energy experiments. Our knowledge on the universal parton distribution functions (PDFs) and fragmentation functions (FFs), which connect the partonic dynamics to the observed hadrons, has been dramatically improved in recent years \cite{Caola:2010cy}.  As a probability density to find a parton (quark or gluon) inside a hadron with the parton carrying the hadron's longitudinal momentum fraction $x$, the PDFs have provided us with the nontrivial and quantitative information about the partonic structure of a hadron.  

In recent years, the hadronic physics community has extended its investigation of partonic structure of hadrons beyond the PDFs by exploring the parton's motion and its spatial distribution in the direction perpendicular to the parent hadron's momentum. Such effort is closely connected to the study and extraction of two new types of parton distributions:  the transverse momentum dependent parton distributions (TMDs)~\cite{Ralston:1979ys,Sivers:1989cc,Kotzinian:1994dv,Mulders:1995dh,Boer:1997nt,Mulders:2000sh,Belitsky:2003nz,Bacchetta:2006tn}, and the generalized parton distributions (GPDs)~\cite{Burkardt:2000za,Goeke:2001tz,Burkardt:2002hr,Belitsky:2003nz,Diehl:2003ny,Ji:2004gf,Belitsky:2005qn,Boffi:2007yc}. 
The ultimate knowledge of finding a single parton inside a hadron --
involving both momentum and space information -- could be encoded in the
phase-space distributions of quantum mechanics, such as the Wigner quasi-probability
distribution $W(k, b)$, whose integration over the parton spatial dependence $(b)$ leads to the TMDs, while its integration over transverse momentum $(k)$ provides the parton's spatial distribution that is relevant to the GPDs. 
A quantum field theory version of the phase-space distributions, 
in terms of the matrix element of the Wigner operator, was discussed in Ref.~\cite{Ji:2003ak}.  Understanding both the momentum and spatial distribution of a parton inside a hadron in terms of the more general Wigner distributions could be the central object of future studies on partonic structure.  Knowledge of TMDs is also crucial for understanding some novel phenomena in high energy hadronic scattering processes, such as, the single transverse spin
asymmetries~\cite{Brodsky:2002cx,Brodsky:2002rv,Collins:2002kn,Ji:2002aa,Belitsky:2002sm,{Boer:2003cm},Kang:2009bp} and small-$x$ saturation phenomena~\cite{Brodsky:2002ue,Iancu:2002xk,Iancu:2003xm,JalilianMarian:2005jf,Marquet:2009ca,Gelis:2010nm,Xiao:2010sa,Xiao:2010sp}. 

\section{Recent theoretical development on TMDs and experimental access}

Like the PDFs, the TMDs and GPDs carry rich information on hadron's partonic structure, while they are not direct physical observables due to the color confinement of QCD dynamics.  It is the leading power QCD collinear factorization theorem \cite{Collins:1989gx} that connects the PDFs to the hadronic cross sections with large momentum transfers: $Q$'s $\gg \Lambda_{\rm QCD}$.  In order to study the TMDs, we need the corresponding TMD factorization theorem for physical observables that are sensitive to parton's transverse motion and the TMDs.  Such observables often involve two very different momentum scales: $Q_1 \gg Q_2 \gtrsim \Lambda_{\rm QCD}$, where the large $Q_1$ is necessary to ensure any perturbative QCD calculation while the small scale $Q_2$ is needed so that these observables are sensitive to the parton's transverse motion.  The transverse momentum distribution of single hadron production in semi-inclusive deep inelastic lepton-hadron scattering (SIDIS) and Drell-Yan lepton pair production in hadronic collisions are two natural examples.  The TMD factorization for these two processes have been carefully examined \cite{Ji:2004wu,Ji:2004xq,Collins:2004nx}.  However, the TMD factorization in QCD is much more restrictive than the leading power collinear factorization. The conventional TMD factorization works for three-types of observables with only two identified hadrons: single hadron $p_T$ distribution in SIDIS, the $p_T$ distribution of Drell-Yan type process, and two-hadron momentum imbalance in $e^+e^-$ collisions. 
But it has been shown to fail for observables with more than two identified hadrons~\cite{Bomhof:2004aw,Collins:2007nk,Vogelsang:2007jk,Collins:2007jp,Rogers:2010dm}.

Important aspects of the TMD parton distributions, such as the gauge invariance, the role of gauge links, and the universality, have been explored in recent years \cite{Collins:2002kn,Ji:2002aa,Belitsky:2002sm,Boer:2003cm,Kang:2009bp,Cherednikov:2007tw,Cherednikov:2008ua,Cherednikov:2009wk}.  Like the PDFs, the definition of TMDs is closely connected to the factorization of physical cross sections, and it is necessary for the TMDs to include all leading power long-distance contributions to the physical cross sections if they could be factorized.  All leading power collinear gluon interactions are summed into the gauge links in the definition of the TMDs.  It is the gauge link that makes the TMDs gauge invariant and provides the necessary phase for generating a sizable single transverse spin asymmetry (SSA) in SIDIS and Drell-Yan processes \cite{Brodsky:2002cx,Brodsky:2002rv,Collins:2002kn,Ji:2002aa,Belitsky:2002sm,{Boer:2003cm}}.  However, unlike the PDFs, which are universal, the TMDs could be process dependent due to the fact that the initial-state and final-state collinear gluon interactions are summed into two different gauge links.  That is, the TMDs extracted from SIDIS could be different from those extracted from Drell-Yan processes because of the difference in gauge links.  Although the TMDs are not in general universal, it could be shown from the parity and time-reversal invariance of QCD dynamics that the process dependence of the spin-averaged as well as spin-dependent TMDs is only a sign, which was referred to as the parity and time-reversal modified universality~\cite{Collins:2002kn,Kang:2009bp}. An important example of the modified universality is that the Sivers function extracted from the SIDIS measurements is opposite in sign from the Sivers function extracted from the Drell-Yan process.  The test of the sign change of the Sivers function from SIDIS to Drell-Yan is a critical test of the TMD factorization.

At leading twist there are eight TMD quark distributions~\cite{Bacchetta:2006tn}: three of them,
the unpolarized, the helicity and the transversity distributions,
 survive in the collinear limit, while the other five vanish in such a limit. 
Each TMD quark distribution explores one unique feature of the quark inside a polarized or unpolarized nucleon.  For example, the Sivers function~\cite{Sivers:1989cc,Sivers:1990fh} provides the number density of unpolarized partons inside a transversely polarized proton, while the Boer-Mulders function~\cite{Boer:1997nt} gives the number density of transversely polarized quarks inside an unpolarized proton.  Although we have gained a lot of information on the collinear PDFs and helicity distributions, we know very little about quark's and gluon's intrinsic transverse motion inside a nucleon. 
Recent measurements of multiplicities and double spin asymmetries as a function of the final transverse momentum of 
pions in SIDIS at JLab \cite{Mkrtchyan:2007sr,Avakian:2010ae} suggest that transverse momentum distributions 
may depend on the polarization of quarks and possibly  also on their flavor. Calculations of transverse momentum 
dependence of TMDs in different models \cite{Lu:2004au,Anselmino:2006yc,Pasquini:2008ax,Bourrely:2010ng} and 
on lattice \cite{Hagler:2009mb,Musch:2010ka} indicate that dependence of  transverse momentum  
distributions on the quark polarization and flavor may be very significant.

Among the TMDs vanishing in the collinear limit, the Sivers function is the best known and has been phenomenologically extracted by several groups mainly from analyzing the azimuthal distribution of a single hadron in SIDIS \cite{Anselmino:2005nn,Collins:2005rq,Vogelsang:2005cs,Anselmino:2008sga}.  However, in the case of positive hadrons, where a signal has been seen,
the measurements of
HERMES~\cite{Airapetian:2009ti} and COMPASS~\cite{Alekseev:2010rw} experiments are only marginally
compatible: the asymmetries measured by COMPASS are somewhat smaller, and
seem to indicate an unexpected dependence on $W$, the mass of the hadronic
final state.
For the transversity distribution,
 there is only one phenomenological extraction by combining the SIDIS and the $e^+e^-$ data \cite{Anselmino:2007fs,{Efremov:2006qm},Anselmino:2008jk}, and  
 information on the rest of the TMDs is rather scarce. Nevertheless, these recent results have already generated great excitement, which is evident from the increasingly active theoretical activities, including modeling and lattice QCD calculations, and planning of future experiments.

A number of experimental facilities, such as COMPASS~\cite{compass-II} at CERN, CEBAF with its 12~GeV upgrade at Jefferson Lab, RHIC at Brookhaven National Lab, Belle at KEK, and in particular, the planned Electron-Ion Collider (EIC), will play a complementary but crucial role in determining these TMD parton distributions.  Among three types of processes where the TMD factorization could be valid, the SIDIS might be the best place to study the TMD parton distributions because of the easy separation of various TMDs with well-determined 
distributions in the azimuthal angles between
the spin, the leptonic plane and the hadronic plane, in addition to the much higher event rates. 
 With a broad energy range and a high luminosity,
the future EIC will be an ideal place to extract the TMDs in a multi-dimensional phase space with a high precision.  Precise measurements of these new distributions could provide us much needed information on the partonic structure of nucleon (nucleus) in order to address the fundamental questions concerning the decomposition of the nucleon spin, and 
the QCD dynamics responsible for the structure of the nucleon. 

The transverse momentum dependence could also be introduced to the hadronization process to get the TMD parton-to-hadron fragmentation functions.  For a quark to fragment into a spinless hadron, such as a pion, there are only two possible fragmentation functions at the leading twist: the unpolarized fragmentation function and the Collins function~\cite{Collins:1992kk}, which is responsible for generating azimuthal asymmetric distribution of hadrons from the hadronization of a transversely polarized quark.  The Collins function has been extracted from recent experiments (Belle~\cite{Abe:2005zx}, HERMES~\cite{Airapetian:2004tw,Airapetian:2010ds}, and COMPASS~\cite{Alexakhin:2005iw,*Alekseev:2008dn}), and was found to be nonzero.  Precise measurements of TMD fragmentation functions provide a new window to explore the dynamics of hadronization.

For cross sections with one large momentum transfer, or several momentum transfers at the same scale, it is more natural and appropriate to use the collinear factorization approach and to expand the cross sections as an inverse power of the large momentum transfer.  Although the leading power term dominates the contribution to the cross sections, it does not contribute to the SSA, which is proportional to the difference of two cross sections with the spin vector reversed.  Like the TMD factorization approach, the SSA in the collinear factorization approach is also generated by the active parton's transverse motion, but, as a net effect after integrating over all possible values of transverse momentum.  Within the collinear factorization approach, the SSA is effectively generated by the quantum interference of two scattering amplitudes: a real amplitude with one active parton and an imaginary part of an amplitude with an active two-parton composite state \cite{Qiu:1991wg,Qiu:1991pp,Qiu:1998ia,Efremov:1981sh,Efremov:1984ip,Kouvaris:2006zy}.  The QCD quantum interference between the amplitude to find a single active parton and that to find a two-parton composite state is represented by a set of new twist-3 three-parton correlation functions.  Unlike the PDFs, which have the probabilistic interpretation of number densities to find a parton within a hadron, these new three-parton correlation functions provides the direct information on the strength of color Lorentz force and/or magnetic force inside a spinning proton.     
The twist-3 contributions are accessible in various spin-azimuthal asymmetries in SIDIS depending on the helicity of the lepton or the hadron. Significant higher-twist asymmetries have been reported by the 
HERMES \cite{Airapetian:1999tv,Airapetian:2006rx,Giordano:2009hi}
and COMPASS Collaborations \cite{Kafer:2008ud} as well as the CLAS and Hall-C Collaborations at 
JLab \cite{Avakian:2003pk,Mkrtchyan:2007sr,Avakian:2010ae}.
 Higher-twist observables, such as longitudinally polarized beam or target SSAs, are important  
for understanding long-range quark-gluon dynamics, and the future EIC due to the wide range in $Q^2$, will 
be an ideal place to pin them down.

Both the TMD factorization approach and the collinear factorization approach at twist-3 provide a viable mechanism to generate the SSA, but, with a very different physical picture.  This is because they cover the SSA in two very different kinematic regimes:  $Q_1\gg Q_2\gtrsim \Lambda_{\rm QCD}$ for the TMD approach while $Q_i\gg \Lambda_{\rm QCD}$ with $i=1,2, ...$ for the twist-3 approach.  Further study has shown that the TMD approach is consistent with the twist-3 approach for the SSA phenomena in a perturbative region, $Q_1\gg Q_2\gg \Lambda_{\rm QCD}$, where they are both valid \cite{Ji:2006ub,Ji:2006vf,Ji:2006br,Bacchetta:2008xw}.  More recently, the evolution equations for the transverse momentum moments of these TMDs have also been investigated, which opens a path for the systematic QCD calculations of SSA beyond the leading order in $\alpha_s$ \cite{Kang:2008ey,Zhou:2008mz,Vogelsang:2009pj,Braun:2009mi}.

Like TMD quark distributions, we could also construct TMD gluon distributions.  But, unlike the quark, gluon does not interact with any colorless particles at the lowest order.  Due to the restriction on the color flow for the TMD factorization to be valid, we only have very few observables that might give the direct access to the TMD gluon distributions, such as the Higgs production at low $p_T$ with an effective $gg\to H^0$ vertex, the momentum imbalance of two isolated photons via an effective $gg\to\gamma\gamma$ vertex, and back-to-back jets or heavy quark pair production~\cite{Mulders:2000sh,Boer:2010zf,Anselmino:2005sh,Meissner:2007rx,ref:xiao_pc} in $ep$ and in $pp$ or $p\bar{p}$ collisions.     

On the other hand, many more observables could access the gluonic sector of twist-3 approach to the SSA.  Heavy flavor production in the DIS regime is a direct probe of gluon content of the colliding hadron or the nucleus.   In particular, the SSAs of open-flavor (anti)D (or B) meson production in the DIS regime provides a unique 
opportunity to measure tri-gluon correlation functions~\cite{Kang:2008qh}, which are closely connected to the gluon's transverse motion and color coherence inside a transversely polarized nucleon. A more recent study shows that there are four tri-gluon correlation functions~\cite{Beppu:2010qn}.
The co-existence of these different tri-gluon correlation functions, which represent the long-distance quantum interference between a gluon state and a two-gluon composite state, is a unique feature of the non-Abelian color interaction.  Motivated by recent calculations~\cite{Kang:2008qh}, preliminary simulations of the event rate and asymmetries for some realistic EIC energies and possible detector coverage have been carried out, which will be presented in the later section of this paper.


Like the PDFs, the TMDs and the GPDs are non-perturbative functions and should be extracted from the experimental measurements of cross sections or asymmetries in terms of relevant factorization formalisms.  In order to get a better picture of the proton's partonic structure from the limited information extracted from the PDFs, the TMDs, and the GPDs, model calculations of these distributions are valuable.  
There have been many interesting model studies recently, see for example~\cite{Schweitzer:2010tt,Avakian:2010br,Boffi:2009sh,Pasquini:2010af,Pasquini:2008ax,Meissner:2007rx}.
These models and their calculations could play a very important role as a first step to describe the experimental observations, to give an intuitive way to connect the physical observables to the partonic dynamics, and to provide key inputs to the partonic structure of the nucleon,  which will help us to address the fundamental questions, such as how the quark spin and its orbital angular momentum contribute to the nucleon spin?

More importantly, very exciting results of TMDs have come from the Lattice
QCD calculations recently~\cite{Hagler:2009mb,{Musch:2009ku},Musch:2010ka}, indicating that
spin-orbit correlations could
change the transverse momentum distributions of partons.
Notable results from Lattice QCD have been obtained for the impact parameter dependent parton distributions, which have a close relation to some interesting TMDs~\cite{Brommel:2007xd}.  With the improvement of computer speed and simulation algorithms, more and more accurate results on the partonic structure from Lattice QCD calculations will become available soon. 

The experimental investigation of multi-dimensional spatial distributions of a parton (or color) inside a bound proton, in terms of the TMDs and the GPDs or the ``mother'' Wigner distributions has just started recently.  Future machines, like EIC, could supply high quality data by scattering polarized leptons off polarized nucleons.  For semi-inclusive reactions, the data with large $Q^2$ and small $P_T$ are dominated by the TMDs, while data with large $Q^2$ and large $P_T$ have the most contributions from pQCD corrections convoluted with collinear PDFs or multiparton correlation functions.  The transition from one regime to the other might be the most interesting aspect and should be carefully studied.  The investigated $x$-region should be as wide as possible to cover both the valence quark region and the unexplored sea quark region.

In summary, while there has been progress on several fronts in the theoretical developments for understanding the transverse momentum dependent parton distributions and fragmentation functions,  
it is just a beginning for us to explore the full picture of partonic structure inside a nucleon.  The future Electron-Ion Collider is a much needed machine to probe the partonic structure of a bound nucleon, to quantify the role of gluons and the color, and to help approach the fundamental question of strong interaction - the confinement of the color.   Given below is a list of questions which have been discussed in various meetings in connection with the TMDs:

\begin{itemize}
\item $Q^2$ evolution.  
The transverse momentum dependence of the
TMDs certainly depends on the large scale $Q$ where the TMDs were
probed. Although the energy evolution equation has been derived for
the TMD
distributions~\cite{Collins:1981uk,Collins:1984kg,Idilbi:2004vb},
very few explicit calculations have been performed to date (e.g.
\cite{Boer:2001he}) to study the $Q^2$ dependence of the associated
experimental observables, such as the azimuthal asymmetries. This
$Q^2$ evolution is not only an important theoretical question, but
also a crucial point to investigate experimentally. An EIC machine
with a wide range of
coverage on $x$ and $Q^2$ for SIDIS processes will provide a great
opportunity to study the scale dependencies of the TMDs in detail.


The transverse momentum distribution of the TMD observables in principle, 
has three characteristic regions: intrinsic, resummation, and perturbative. 
In practice, a Gaussian transverse momentum distribution has been used to 
fit the existing experimental data in order to extract the TMDs. At low 
collision energy, there is not much phase space for the gluon shower around 
the hard collision, the active parton's intrinsic $p_T$ distribution dominates,
and therefore, a Gaussian distribution should be a good approximation.
However, 
with a much higher collision energy at an EIC machine and a larger phase space
for the gluon shower, the Gaussian distribution will not be adequate to 
describe the observed $p_{T}$ distribution in SIDIS. Instead, a distribution with
a proper resummation of large logarithmic contributions from the gluon
shower should be used~\cite{Nadolsky:1999kb,Qiu:2000ga,Qiu:2000hf}.
When $p_T$ is as large as the hard scale $Q$, a perturbative 
calculated $p_T$ should be more relevant~\cite{Qiu:2000ga,Qiu:2000hf}. Investigation of the resummation
or ``matching'' region, especially as a function of $Q^2$, will provide an 
important test of the theoretical framework, i.e.\ TMD factorization.

\item Relation of TMDs to the parton orbital angular momentum. There have
been qualitative suggestions about the connection of the quark orbital angular momentum 
to the TMDs~\cite{She:2009jq,Pasquini:2008ax,Avakian:2010br}. However, we still
do not have a rigorous way to build this connection. Certainly, 
model calculations will help shed light on this important issue.

Besides the connection to the parton's orbital angular momentum, the TMDs should provide much richer information on nucleon structure in momentum space. We need more theoretical
investigations along this direction.  

\item Global study at next-to-leading order. So far, all phenomenological
studies are limited to the leading order in perturbative QCD. We have to
go beyond this simple picture to build a systematic framework to extract
the TMDs.

\item Small-$x$ parton distributions. The investigation of TMDs at
small-$x$ (sea) has started, and progresses have been made
recently~\cite{Marquet:2009ca,Xiao:2010sp,Dominguez:2010xd,Dominguez:2011wm}.
In particular, it was found that di-hadron/dijet correlation
in DIS processes in $eA$ collisions can be used to probe the
Weizacker-Williams gluon distribution formulated in the
color-glass-condensate
formalism~\cite{Dominguez:2010xd,Dominguez:2011wm}. However, more
theoretical studies are needed to build a rigorous connection
between the TMD and CGC approaches.

\item Universality of the TMDs.  Much of the predictive power of QCD factorization relies on the universality of nonperturbative distributions.  More work is needed to better understand the process dependence of the TMDs and their connections to what Lattice QCD can calculate, which is crucial for the predictive power of the TMD factorization and physical interpretation of the TMDs.

\end{itemize}

We believe that the TMD community as it addresses the above and other important questions will naturally make the case for an EIC stronger and will be ready for the new era of QCD and hadron structure.
In the next section, we will present simulations that have been carried out with a goal of addressing the aforementioned questions with a high luminosity electron-ion collider.

\section{SIDIS at an EIC}

\subsection{Kinematics}
In an EIC, a beam of electrons collides with a beam of ions. 
The SIDIS process requires to detect both the scattered electron and one of the leading 
hadrons produced in the final state. In general, the process can be expressed as:
\begin{equation}\label{SIDIS}
  \ell(P_e^i) + N(P) \to \ell'(P_e) + h(P_h) + X
\end{equation}
where $\ell$, $N$,  $\ell'$ and $h$ denote the initial electron, the initial 
proton~\footnote{We assume that the ion is proton for simplicity. In principle, the same argument also applies to any ion beam.}, the 
scattered electron, and the produced hadron in the final state, respectively. All the
four-momenta are given in parentheses. 

Under the one-photon exchange approximation, the four-momentum of the virtual photon 
is expressed as $q = P_e^i - P_e$ and the four momentum transfer square is $q^2 = - Q^2$.
The relevant Lorentz invariant variables are defined as:
\begin{eqnarray} \label{eqn:xyz}
x = \frac{Q^2}{2 P\cdot q},
y = \frac{P \cdot q}{P \cdot P_e^i},
z = \frac{P \cdot P_h}{P\cdot q},
s = (P_e^i +P)^2 \approx 4 E_e^i \cdot E_P.
\end{eqnarray}
Here, $x$, also referred to as Bjorken $x$, represents
the initial nucleon momentum fraction carried by the parton in the infinite momentum frame, $y$ and $z$ are the fractional momentum carried by the virtual photon and the leading hadron, respectively, and $s$ is the center-of-mass energy squared of the initial electron-nucleon system. The last 
approximation in Eq.~\ref{eqn:xyz} is made by neglecting the masses of the electron and the nucleon,
which are much smaller than the center-of-mass energy at EIC kinematics. 

With approximations, one can immediately obtain 
\begin{eqnarray} \label{eqn:Q2xys}
Q^2 = x \cdot y \cdot s,
\end{eqnarray}
which clearly illustrates the relation between $x$ and $Q^2$ at fixed $s$.

\begin{figure}[tbp]
\centering
\includegraphics[width=10cm]{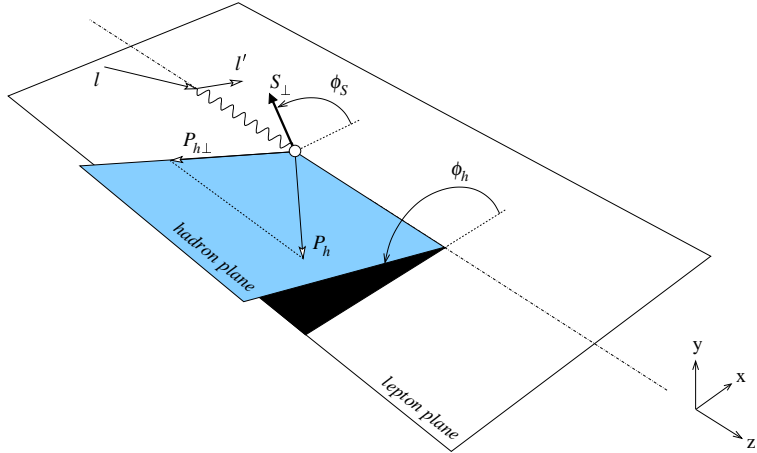}
\caption{Definitions of azimuthal angles $\phi_h$ and $\phi_S$, and the hadron
  transverse momentum for SIDIS in the ion-at-rest frame~\cite{Bacchetta:2006tn}.}
\label{fig:anglestrento}
\end{figure}

In addition to the aforementioned Lorentz invariant variables, there are a few 
frame-dependent kinematic variables, $\phi_S$, $\phi_h$, and $P_{T}$ (the target spin angle, 
the azimuthal angle and the transverse momentum of the leading hadron), 
which are also essential to SIDIS process. They are defined according to the Trento
convention as illustrated in Fig.~\ref{fig:anglestrento} in the nucleon-at-rest frame~\footnote{More 
generally, the $\phi_S$, $\phi_h$, and $P_{T}$ are defined in the collinear frame, where the virtual
photon moves col-linearly with the initial nucleon. The nucleon-at-rest frame is a special
situation of the collinear frame.}.

\subsection{Phase Space Coverage}~\label{sec:phase}

\begin{figure}
  \begin{center}
    \begin{minipage}[t]{0.48\linewidth}
      \epsfig{file=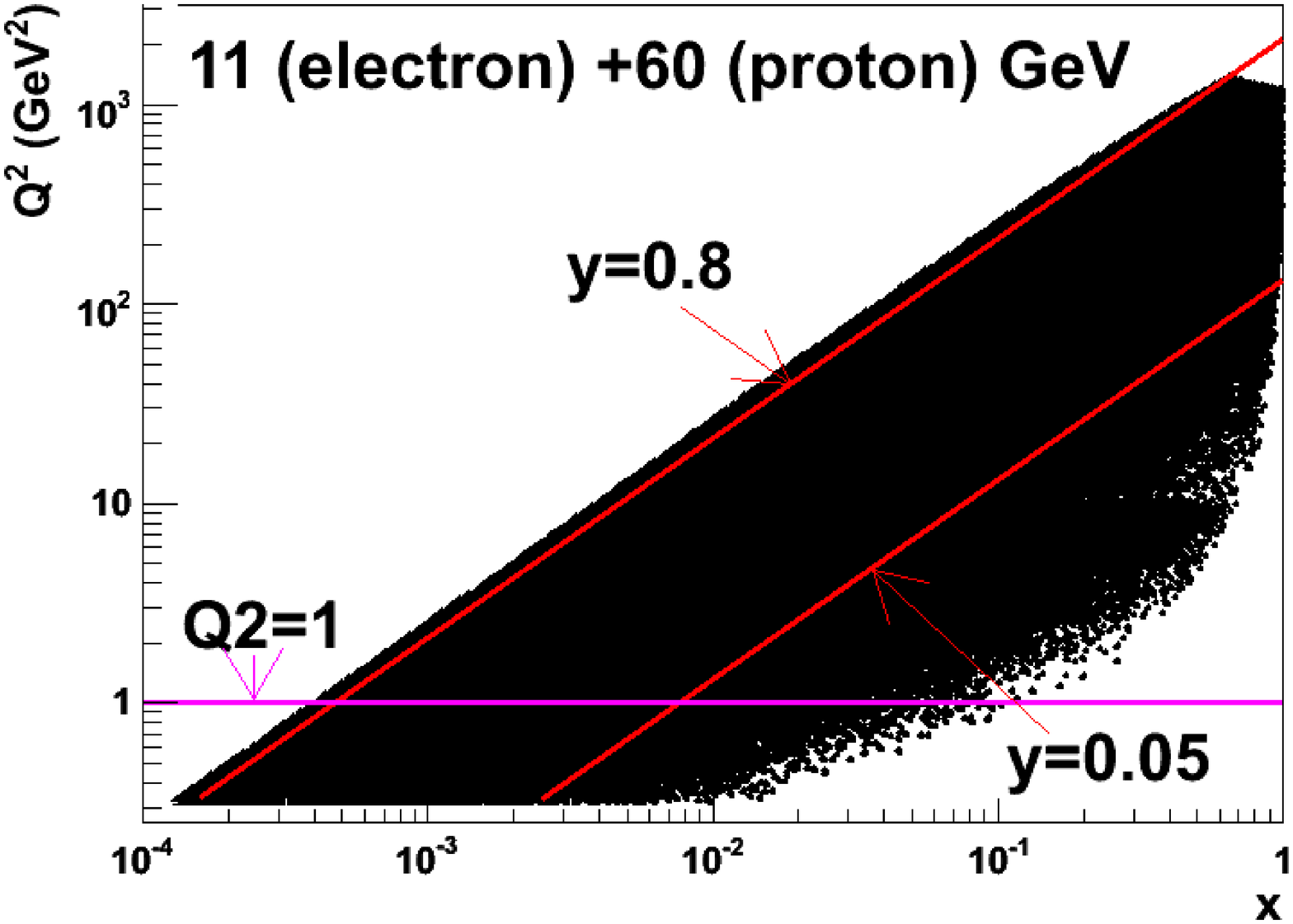, width=\linewidth}
      \caption{Phase space of Q$^2$ vs. $x$ with DIS and $y$ cuts illustrated. No SIDIS
      cross section is applied. The ``p'' in the legend refers to the fact that the ion beam is the proton beam.}
      \label{fig:whole}
    \end{minipage}\hfill
    \begin{minipage}[t]{0.48\linewidth}
      \epsfig{file=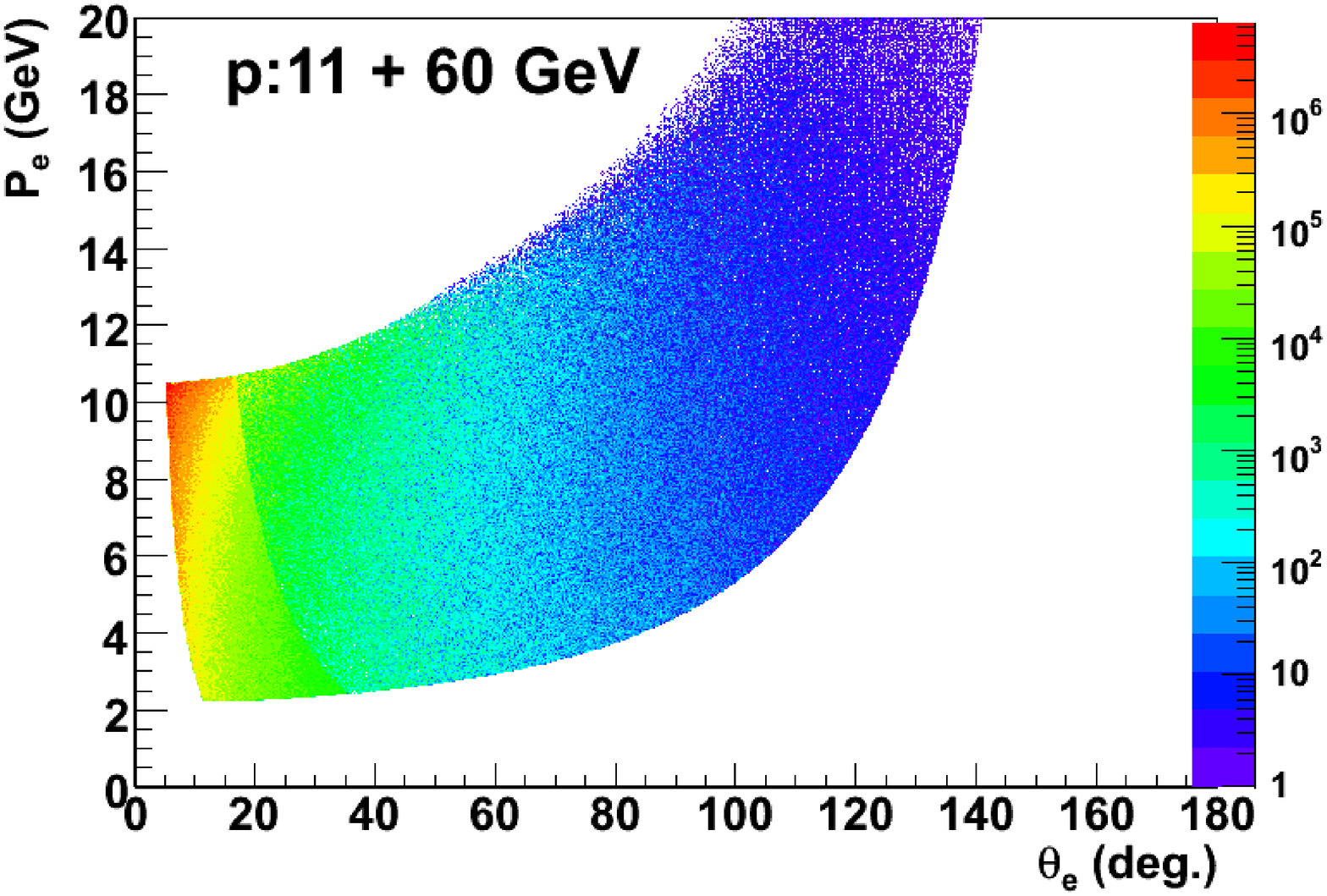, width=\linewidth}
      \caption{Momentum vs. polar angle in the lab frame for the scattered electron after weighting
	by the SIDIS differential cross sections. Here, $0^{\circ}$
represents the momentum direction of the initial electron beam.}
      \label{fig:elepth}
    \end{minipage}
  \end{center}
\end{figure}

In this section, we discuss the SIDIS phase space coverage mainly with 
the 11+60 GeV configuration, which represents a 11 GeV electron beam colliding
with a 60 GeV proton beam. In the simulation, the scattered electrons are generated in momentum
P$_e > 0.7$ GeV/c, polar angle $2.5\,^{\circ} < \theta_e < 150 \,^{\circ}$ and full azimuthal angle. 
Fig.~\ref{fig:whole} shows the Q$^2$ vs. $x$ phase space for the 11+60 GeV configuration. Since we 
are mainly interested in the DIS region, the following cuts, $Q^2>1$ GeV$^2$ and $W>2.3$ GeV, are applied. In addition, the 
$0.05< y < 0.8$ cut is also applied. Here, the 0.8 cut-off is chosen to reflect the lowest
detectable energy of the scattered electrons, which is usually limited by the hardware acceptance and
uncertainties in the radiative correction~\footnote{The larger the $y$ value is, the more radiative 
  correction should be applied, which would lead to larger systematic uncertainties.}. The 0.05
cut-off of $y$ is limited by the resolution of $x$. As shown in the following equation:
\begin{eqnarray}\label{eqn:deltay}
\frac{\delta x}{x} & =&  \frac{\delta P_e}{P_e} \cdot (\frac{1}{y}) + \frac{\delta
  \theta_e}{\tan{\frac{\theta_e}{2}}} \cdot (1 + \tan^2{\frac{\theta_e^f}{2}} \cdot (1-\frac{1}{y}) ) \\\nonumber
& \approx & \frac{\delta P_e}{P_e} \cdot (\frac{1}{y}) + 2 \frac{\delta
  \theta_e}{\theta_e},
\end{eqnarray}
assuming a fixed momentum resolution $\delta P_e/P_e$, the resolution of $x$ increases 
dramatically at small $y$. Therefore, the $y>0.05$ cut
is applied in the simulation in order to maintain a reasonable $x$ resolution for forward
electron detection. With the above cuts applied, Fig.~\ref{fig:elepth} shows the distribution of 
momentum vs. polar angle of the scattered electron in the lab frame after weighting
each event by the SIDIS differential cross section. Two observations are made to reflect 
the needs in the detector design: 
\begin{itemize}
\item Most of the scattered electron are concentrated in the high-momentum region (closer to the initial 
electron momentum), which is corresponding to the small $y$ region. This is consistent with the fact that SIDIS
cross section are larger at smaller $y$.
\item No electrons are distributed at very forward angles ($\le 5^\circ$) due to  
  the $Q^2>1$ GeV$^2$ cut applied. There is no essential need to detect very forward-angle electrons.
\end{itemize}

\begin{figure}
  \begin{center}
      \epsfig{file=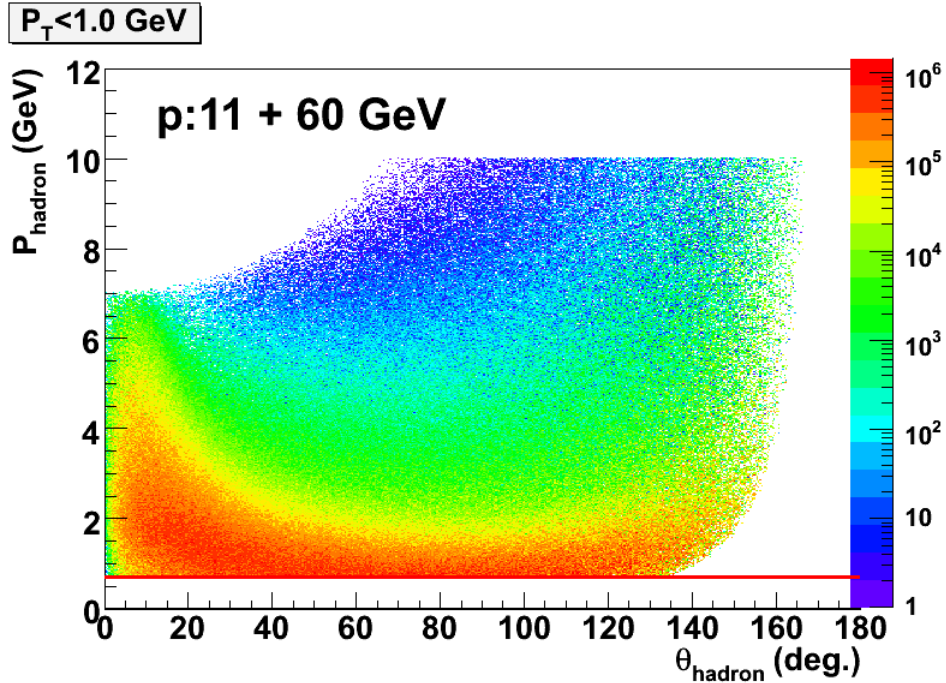, width=5.0in}
      \caption{Momenta vs. polar angles for the detected hadron in the lab frame
	(weighted by differential cross section). The 
  180$^\circ$ represents the initial momentum direction of the ion beam.}
      \label{fig:hadpth}
  \end{center}
\end{figure}

For SIDIS process, more cuts are applied on the hadron side. They are $0.2< z < 0.8$ and 
$M_X>1.6$ GeV cut, where $M_X$ is the missing mass of the $X$ system in Eq.~\ref{SIDIS}. 
The 0.8 cut-off in $z$ excludes the events from exclusive channels. The 0.2 
cut-off is required to stay in the current fragmentation region, where the detected hadron
can be used to tag the struck quark. In addition, we also apply a low $P_T$ cut ($P_T<1$ GeV/c)
for the TMD physics, and a $P_T>1$ GeV/c cut for the large $P_T$ physics. 
Fig.~\ref{fig:hadpth} shows the momenta of detected hadrons vs. polar
angles in the lab frame. Events are  weighted by the SIDIS differential cross section.
In this simulation, the hadrons are generated for 0.7 GeV/c $< P_{hadron} <$ 10 GeV/c, full
polar and azimuthal angular coverages. The  $P_T<1$ GeV/c cut is applied. Three observations
are made to reflect the needs in the detector design:
\begin{itemize}
\item Most of the hadron events are concentrated in the momentum region of 0.7-7 GeV/c. There is
no essential need to cover very high momentum region.
\item The hadrons have a wide distribution of the polar angle in the lab frame. 
\item No essential need to cover the very backward angle for the hadron lab polar angle. However, 
a large backward angular coverage is important for the study of SSA from SIDIS in the target 
fragmentation region~\cite{Peng:private}. 
\end{itemize}

\begin{figure}
  \begin{center}
    \begin{minipage}[t]{0.48\linewidth}
      \epsfig{file=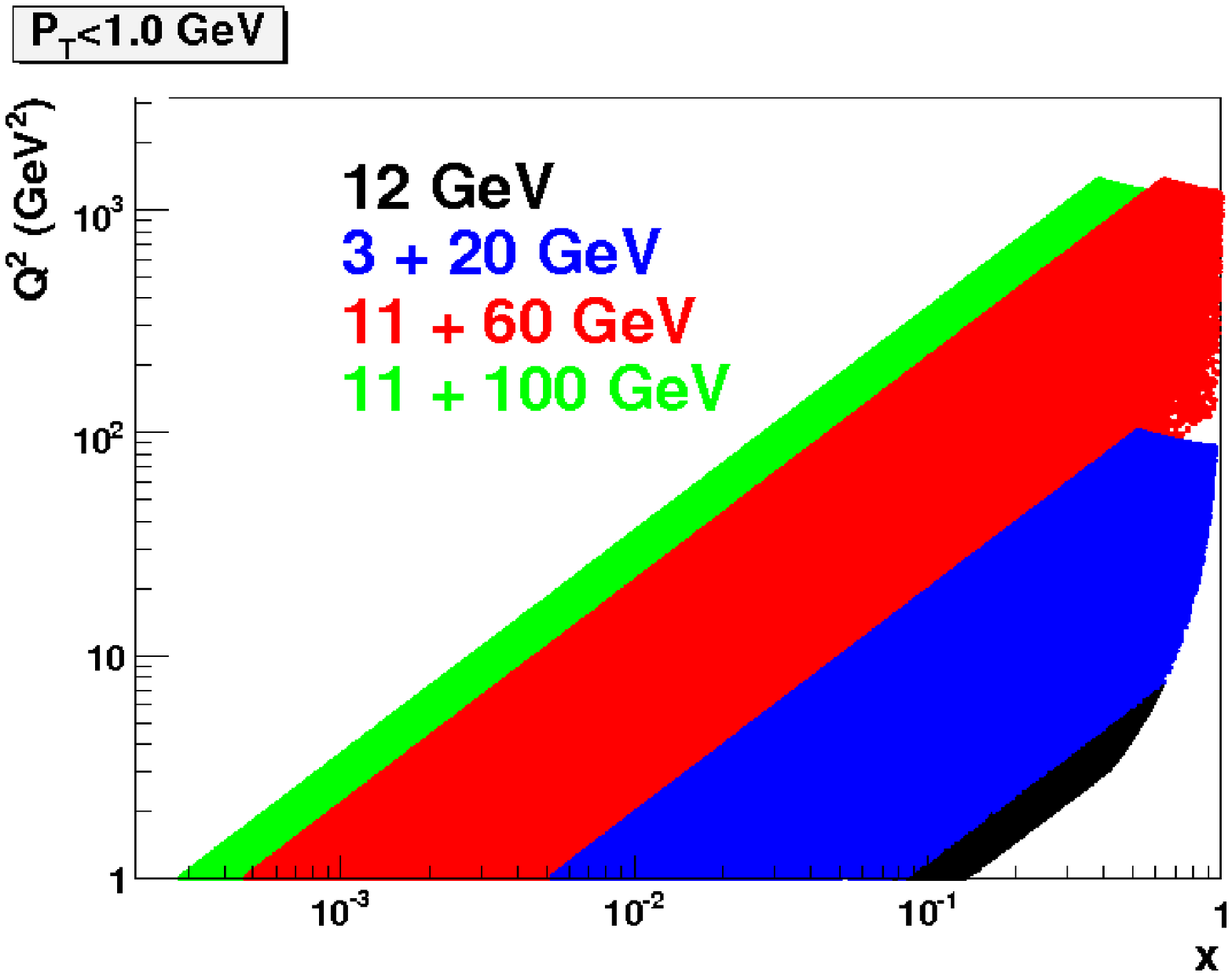, width=\linewidth}
      \caption{Mapping of SIDIS phase space of different energy configurations with
      proton beam. The 12 GeV phase space is shown in the black band. The blue, red, green bands
      represent the phase space of 3+20, 11+60 and 11+100 GeV configurations, respectively.}
      \label{fig:energies}
    \end{minipage}\hfill
    \begin{minipage}[t]{0.48\linewidth}
      \epsfig{file=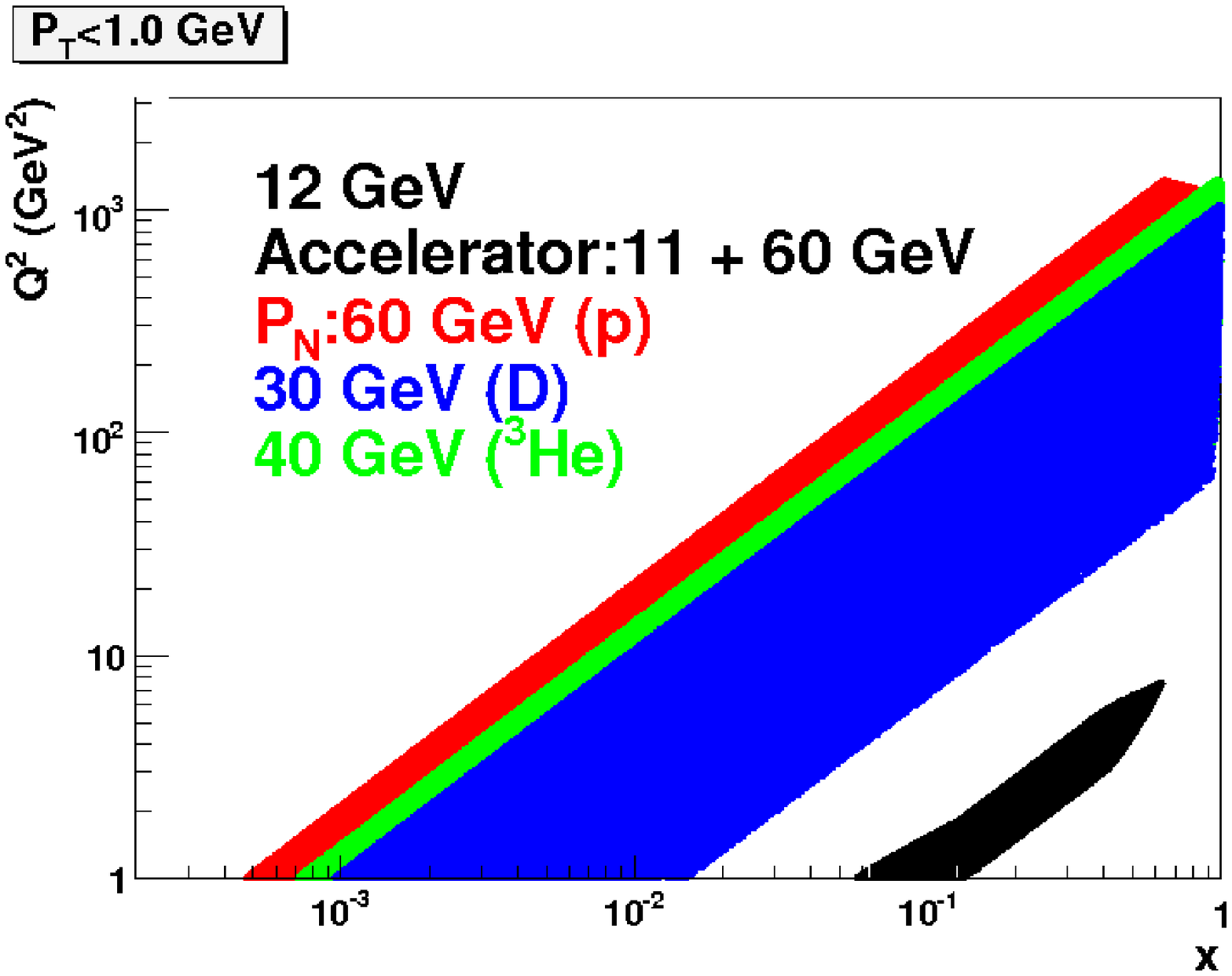, width=\linewidth}
      \caption{Mapping of SIDIS phase space of different ion beams, given the 
	fixed accelerator configuration.}
      \label{fig:ions}
    \end{minipage}
  \end{center}
\end{figure}

The upcoming JLab 12 GeV upgrade would access the SIDIS phase space at low $Q^2$ and high $x$ region
due to the smaller $s$~\cite{12GeV_transversity,Gao:2010av,CLAS_TMD1,CLAS_TMD2,CLAS_TMD3,CLAS_TMD4,TMD_SBS}. The black band in Fig.~\ref{fig:energies} shows the phase space of the approved 
11-GeV SoLID SIDIS experiment~\cite{Gao:2010av,12GeV_transversity}. In order to bridge between the phase spaces of
11+60 GeV configuration and the JLab 12 GeV upgrade, a low-energy configuration of EIC, e.g. 3+20 GeV configuration 
is strongly desired. Such a configuration would overlap with both phase spaces of the 11+60 GeV configuration 
and the JLab 11-GeV fixed-target experiment. In addition, a higher-energy configuration, 11+100 GeV (green
band in Fig.~\ref{fig:energies}), would extend the study of SIDIS process to even lower $x$ and higher $Q^2$ 
regions.

In order to achieve a quark flavor separation from the SIDIS data, measurements with both proton and neutron
are essential. Since there is no free high-energy, high-intensity neutron beam available, one has to use the light ion beam instead.
A deuteron beam is a natural choice. In the polarized
case, the $^3$He ion has a unique advantage as the 
effective polarized neutron beam; the ground state of $^3$He is dominated by the S-state, 
where the two protons are arranged with spin anti-parallel to each other. Therefore, the $^3$He spin is 
dominated by the neutron spin. However, the phase space of the ion is not the same as
that of the proton. Given a fixed accelerator configuration, momentum per nucleon in
an ion is proportional to $Z/A$, in which $Z$ is the atomic number,
and $A$ is the mass number. Therefore, the light ion beam would lead to a smaller $s$ with the same accelerator
configuration.  Fig.~\ref{fig:ions} illustrates
different mapping of these three ion beams (accelerator: 11+60 GeV configuration~\footnote{60 GeV represents the 
momentum for proton.}). Therefore,
\begin{itemize}
\item The lowest achievable $x$ value for quark flavor separation is limited by the light ion beam rather than the 
  proton beam.
\item The highest achievable $Q^2$ value for quark flavor separation is also limited by the light ion beam.
\end{itemize}

\begin{figure}[tbp]
\centering
\epsfig{file=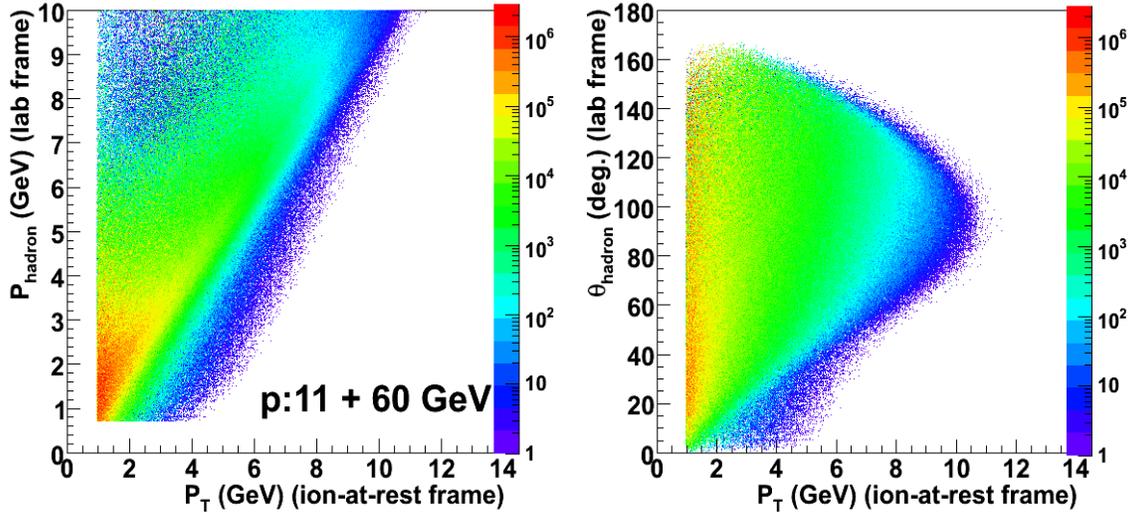,height=3.in,width=6.in}
\caption{The detected momenta of hadrons  vs. P$_T$ (left) 
  and polar angles vs. $P_T$ (right) at $P_T>$1 GeV/c.}
\label{fig:hpt_pth_had}
\end{figure}

At high $P_T$ region ($P_T>1$ GeV/c), the requirements on the hadron detection 
are shown in Fig.~\ref{fig:hpt_pth_had}. The momenta of the hadron (left) and the lab polar angles of the hadron (right) are plotted vs. $P_T$. We make the following observations:
\begin{itemize}
\item The hadron momentum range will increase with the increment of $P_T$.
\item The hadron lab angles distribute widely over the entire phase space.
\item It is not essential to cover the very backward angular range for the hadron lab polar angle.
\end{itemize}

\subsection{Transverse Single Spin Asymmetry Measurements for Light Mesons}

At an EIC, the transverse single spin asymmetry (TSSA) measurements with an unpolarized electron
beam and a transversely polarized proton (or effective neutron) beam can provide rich information
on the transverse spin structure of the nucleon. Three leading twist TMDs, transversity, Sivers and 
pretzelosity distributions can be accessed. A large $Q^2$ coverage 
of EIC will allow for a detailed study of the $Q^2$ evolution of the TMDs. The coverage in the small $x$ region is essential to study sea quark TMDs. In particular, the 
light-meson ($\pi^{\pm}$, $K^{\pm}$) SIDIS process will allow for a map of the TSSA for the sea quarks at low 
$Q^2$ and for the valence quarks at high $Q^2$. Since TSSAs of SIDIS depend on four 
kinematic variables $x$, $z$, $Q^2$, and $P_T$, a complete understanding would require mapping the TSSA 
in 4-D phase space. Therefore, a high luminosity machine is essential. In this section, we will illustrate 
the impact of a high luminosity EIC on TSSA measurements.

\subsubsection{Monte-Carlo Method}
Since most of the TSSAs are relatively small~\footnote{As shown later, the 
expected asymmetries on proton is about a few percent, while the asymmetries on light 
ion are even smaller due to dilutions from spectator nucleon(s).}, the projected uncertainty of the measured 
asymmetries can be approximated as:
\begin{eqnarray}\label{Eq:asymmetry}
\delta A_{N} & = & \frac{1}{P_{e} P_{I} P_{N} f_{D}} \cdot \frac{1}{\sqrt{N_{raw}}} \cdot \sqrt{1-A^2} \nonumber \\
& \approx &  \frac{1}{P_{e} P_{I} P_{N} f_{D}} \cdot \frac{1}{\sqrt{N_{raw}}}
\end{eqnarray}
where $P_e$, $P_I$ and $P_N$ are the polarizations of the electron, ion beam, 
and effective polarization of the nucleon. The $f_{D}$ 
is the effective dilution factor, and $N_{raw}$ is the raw measured counts summing over the 
two spin states. In the case of a proton beam, $P_{N} = 1$ and $f_{D} = 1$. In the case of 
a $^3$He beam, $P_{N} = 87.5\%$ and $f_{D} \sim 0.3 $. 

In addition, $N_{raw}$ is the measured counts summing over the two spin states. Therefore, it is only
proportional to the unpolarized cross section. The following Monte-Carlo procedure is adopted
to obtain the projections on the separated Collins, Sivers and pretzelosity asymmetries. 
\begin{itemize}
\item Simulate scattered electrons and generate pions/kaons uniformly in both momentum and 
  coordinate space in the lab frame. Cuts are applied to mimic the expected 
  detector acceptance.
\item Apply the SIDIS cuts as described in Sec.~\ref{sec:phase}.
\item Calculate the SIDIS differential cross section for each accepted event. In this 
  step, one has to calculate a 6x6 Jacobian matrix (see Sec 9.7 of Ref.~\cite{xqian_thesis} 
  for the complete derivation) to transform the SIDIS differential cross 
  section to the lab frame. 
\item Combining with the expected luminosity, running time, one can calculate the expected raw 
  number of events in each of the 4-D kinematic bin.
\item The projected uncertainties on the raw asymmetry are obtained after including the
  beam polarizations, nucleon effective polarization and the effective dilution factor. 
\item Additional factors are introduced to mimic the increase of uncertainties due to the 
azimuthal angular separation of Collins, Sivers, and pretzelosity asymmetries. The detailed 
discussion of these factors can be found in Appendix II of Ref.~\cite{12GeV_transversity}.
In the case of a full and uniform azimuthal angular coverage of $\phi_H$ and $\phi_S$, these
factors equal to $\sqrt{2}$, and are independent of the number of terms used in the 
fitting. More generally, they depends not only on the angular coverage, but also on the 
event distribution of the azimuthal angle. The factor on the Collins asymmetry is the same
as that of the pretzelosity asymmetry, and slightly different from that of the Sivers asymmetry. 
In practice, these factors are calculated based on the simulated event distribution from 
Monte-Carlo.
\end{itemize}

\subsubsection{\label{sec:dxs}Calculation of SIDIS Differential Cross Section}
In this section, we provide more details in how the SIDIS differential cross sections 
are calculated. The cross section at small $P_T$ (P$_T<$1 GeV/c)  is calculated based on
the TMD formalism~\cite{Bacchetta:2006tn}. 
\begin{eqnarray} \label{eqn:crossmaster}
\frac{d\sigma}{d\xbj \, dy\, d\psi \,dz\, d\phi_h\, d P_{T}^2}
= \frac{\alpha^2}{\xbj y\slim Q^2}\,
\frac{y^2}{2\,(1-\varepsilon)}\,  \biggl( 1+\frac{\gamma^2}{2\xbj} \biggr)\,
F_{UU ,T},
\end{eqnarray}
where 
\begin{eqnarray} \label{eqn:s}
\gamma = \frac{2 M x}{Q},
\end{eqnarray}
and $\alpha$ is the fine structure constant.  The angle $\psi$ is the 
azimuthal angle of $\ell'$ around the lepton beam axis with respect to an arbitrarily fixed
direction, which in case of a transversely polarized ion, it is chosen
to be the direction of $\vec{S}$.  The corresponding relation between
$\psi$ and $\phi_S$ is given in Ref.~\cite{Diehl:2005pc}; in deep
inelastic kinematics one has $\de \psi \approx \de \phi_S$. 
The structure function, $F_{UU ,T}$ on the r.h.s. can be expressed as:
\begin{eqnarray}
F_{UU,T}
& = & \xbj\,\sum_a e_a^2 \int
d^2 \bpt \,f(x,\bpt^2)\,\hF(z,\vert \bP_{T}-z \bpt\vert^2),
\end{eqnarray} 
which depends on $x$, $Q^2$, $z$ and $P_{T}^2$.
Here, the first and second subscript of the above structure function indicate
the respective polarization of the lepton and the ion beam, whereas the third
subscript ``T'' specifies the polarization of the virtual
photon with respect to the virtual photon momentum direction. 
The conversion to the experimentally relevant longitudinal or transverse 
polarization w.r.t.\ the lepton beam direction is straightforward and given in
\cite{Diehl:2005pc}.  The ratio $\varepsilon$ of longitudinal and
transverse virtual photon flux in Eq.~\ref{eqn:crossmaster} is given by
\begin{eqnarray}\label{eqn:epsilon}
\varepsilon = \frac{1-y -\frac{1}{4}\slim \gamma^2 y^2}{1-y
  +\frac{1}{2}\slim y^2 +\frac{1}{4}\slim \gamma^2 y^2} ,
\end{eqnarray}

In order to calculate $F_{UU,T}$, the Gaussian ansatz~\cite{Bacchetta:2006tn,Diehl:2005pc},
is adopted for the transverse momentum dependent parton distribution $f(x,\bpt^2)$
(TMD) and fragmentation function $D(x,\bkt^2)$ (FF):
\begin{eqnarray}
f(x,\bpt^2) & = & f(x,0)\,\exp (-R_H^2 \bpt^2) \\
\hF(z,\bkt^{2}) & = & \hF(z,0)\,\exp (-R_h^2 \bkt^2).
\end{eqnarray}
Therefore, $F_{UU,T}$ becomes
\begin{eqnarray}
F_{UU,T} & = & \xbj\,\sum_a e_a^2 f(x)\,\hF(z)\,\frac{{\cal G}(Q_T;R)}{z^2},
\end{eqnarray}
where ${\cal G}(Q_T;R)$ = $(R^2/\pi)\,\exp(-Q_T^2R^2)$, i.e., a
Gaussian of which $Q_T=P_{T}/z$, and the fall-off is determined by a radius $R$. 
Such radius is related to the radii $R_H$ and $R_h$ governing the fall-off of
$f(x,\bpt^2)$ and $D(x,\bkt^2)$ as $R^2$ = $R_H^2\,R_h^2/(R_H^2 + R_h^2)$.
Furthermore, the CTEQ6M~\cite{Pumplin:2002vw} is used to parametrize 
the parton distribution function (PDF) $f(x,0)$. The parametrization of the unpolarized
fragmentation function $\hF(z,0)$ is from Ref.~\cite{Binnewies:1995pt}. $R_H^2$ and $R_h^2$
are assumed to be 0.25 GeV$^2$ and 0.2 GeV$^2$~\cite{Anselmino:2005nn}, respectively.
While the assumption of the $x$ and $\bpt^2$ factorization in addition to the Gaussian ansatz for the 
$\bpt^2$ dependence has been used widely in the literature, a statistical model for TMDs~\cite{Bourrely:2010ng} 
has been developed recently which involves a non factorisable $x$ and $\bpt^2$ dependence, and the comparison with
those which also have non factorisable TMDs, based on the relativistic covariant 
method~\cite{Zavada:2007ww,Efremov:2009vb,Efremov:2010cy} 
has been made. We remark that in no TMD model considered so far~\cite{Jakob:1997wg,Gamberg:2007wm,Courtoy:2008vi,Courtoy:2008dn,Courtoy:2009pc,
Avakian:2008dz,Pasquini:2008ax,
Bacchetta:2008af,Bacchetta:2010si,Meissner:2007rx,Avakian:2010br,Boffi:2009sh,Pasquini:2010af,
Ellis:2008in}
  the Gaussian Ansatz is strictly supported, although some
  support it approximately~\cite{Avakian:2010br} and so does phenomenology~\cite{Schweitzer:2010tt}.

\begin{figure}[tbp]
\centering
\includegraphics[width=0.65\textwidth]{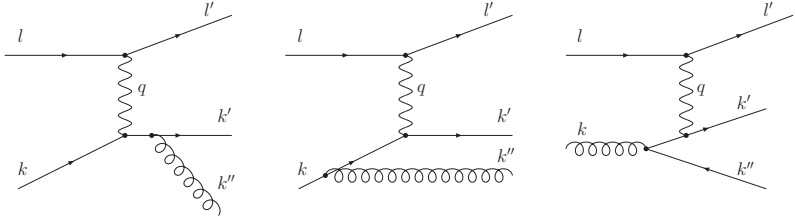}
\caption{Feynman diagrams corresponding to the NLO contribution at high P$_T$ region.}
\label{fig:feynman_largept}
\end{figure}

The cross section at large $P_T$ (P$_T>$1 GeV/c) is dominated by the pQCD higher-order collinear 
contributions~\cite{Anselmino:2006rv}. The dominant partonic processes are shown by the Feynman diagrams
in Fig.~\ref{fig:feynman_largept}. The quark can emit a hard gluon, or be generated by the gluon through pair production. The expression of the $\mathcal{O}(\alpha_s)$ (LO) calculation of the differential cross section in the high P$_T$ region can be found in Ref.~\cite{Anselmino:2006rv}, in which 
the $R_H^2$ and $R_h^2$ are assumed to be 0.28 GeV$^2$ and 0.25 GeV$^2$~\cite{Anselmino:2006rv}, respectively.

The cross sections in both low and high $P_T$ regions are 
expressed as $\frac{d\sigma}{d\xbj \, dy\, d\psi \,dz\, d\phi_h\, d P_{T}^2}$ in 
the ion-at-rest frame. A 6$\times$6 Jacobian matrix~\cite{xqian_thesis} is derived to convert the
calculated differential cross section to the lab frame in terms of the kinematic variables 
of the final-state lepton and hadron: $P_{e}$, $\theta_{e}$, $\phi_{e}$, $P_{h}$,
$\theta_{h}$, $\phi_{h}$.

\begin{figure}[tbp]
\centering
\epsfig{file=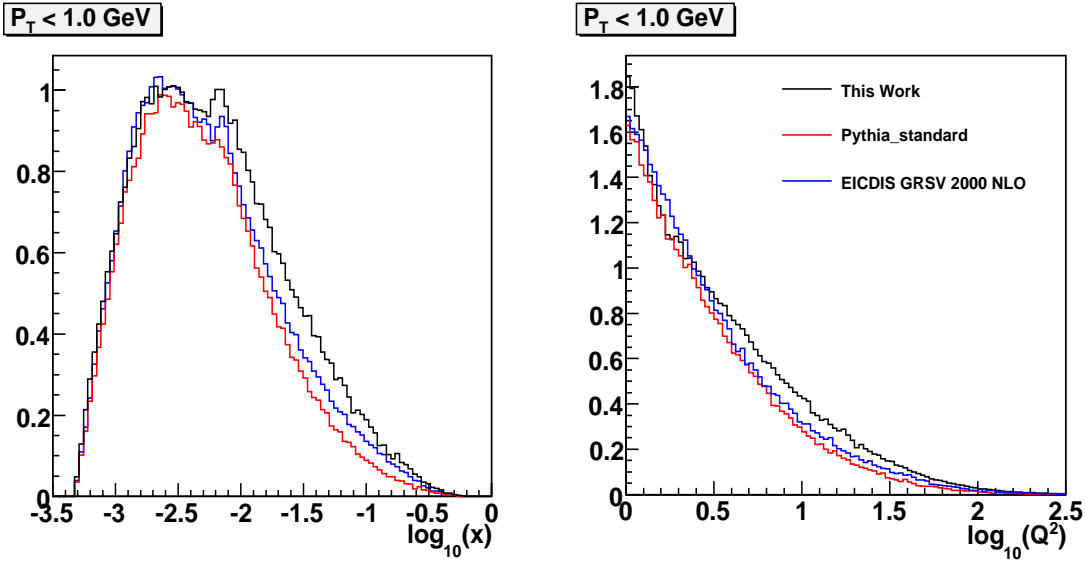,height=3.in,width=6.in}
\caption{The SIDIS comparisons at $P_T<$1 GeV/c for charged pion electroproduction. The EIC configuration is 10+60 GeV.}
\label{fig:x_Q2_lpt_comp}
\end{figure}

\begin{figure}[tbp]
\centering
\epsfig{file=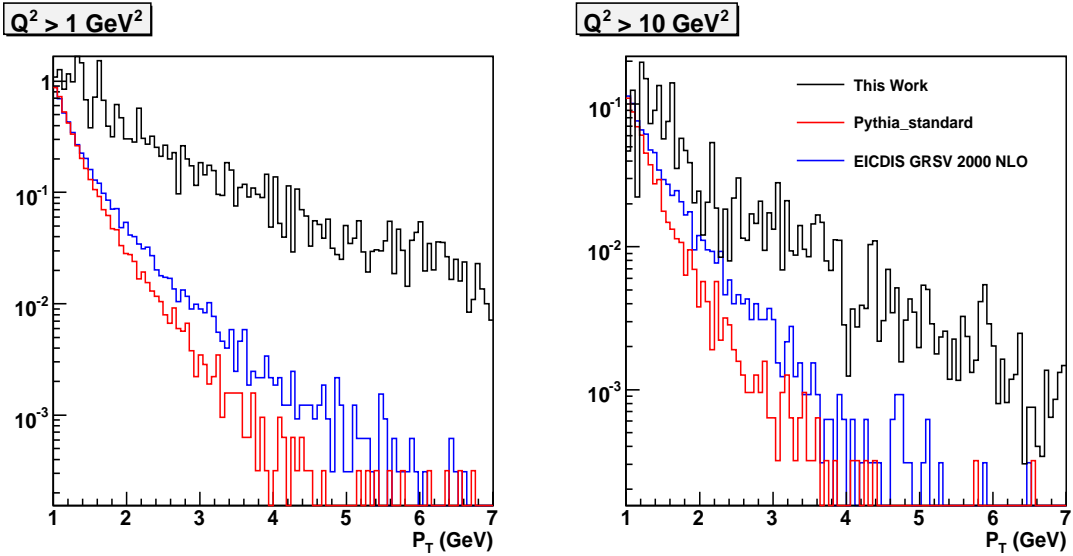,height=3.in,width=6.in}
\caption{The SIDIS comparisons at Q$^2>$1 GeV$^2$ and Q$^2>$10 GeV$^2$.}
\label{fig:pt_comp}
\end{figure}

We have carried out studies to compare results from our approach described above with those from EICDIS~\cite{ref:harut_pc}, which is based on 
PEPSI generator, and results from PYTHIA generator~\cite{ref:elke_pc}. 
The PDF parametrization used in EICDIS is according to GRSV2000 NLO model~\cite{Gluck:2000dy}, while the input of PYTHIA is the standard scenario~\cite{ref:elke_pc}. 
We compare the SIDIS process for charged pion production between our results and those from the two models mentioned above for the 11+60 GeV EIC configuration in Fig.~\ref{fig:x_Q2_lpt_comp} and Fig.~~\ref{fig:pt_comp}.
In this comparison, the fragmentation function by de Florian, Sassot, and Stratmann~\cite{deFlorian:2007aj} is used.
 While  the shape in the
cross sections we obtain is in good agreement with those from the two models in the low $P_T$ region, the results 
from the two models have been scaled down by a factor of 1.5 shown in Fig.~\ref{fig:x_Q2_lpt_comp}.
Our results are about 1.5 lower than those from EICDIS and PYTHIA. 
Such a difference is likely due to missing contributions of longitudinal polarized photon as well as diffractive processes in our approach. 
Therefore, we take the conservative approach of using our rates for 
projections. 
On the other hand as shown in Fig.~\ref{fig:pt_comp}, there are considerable differences between our results and those from the two models in the high $P_T$ region. However, the differences become smaller in the higher Q$^2$ region (Q$^2>$10 GeV$^2$). While more studies are needed in order to understand these differences, we decided to use correction factors to adjust the distributions from our approach to match those from the two models in order to make projections in the high $P_T$ region (Fig.~15). At this stage of the study, this temporary solution is probably adequate. An EIC machine will naturally address 
this $P_T$ dependence and provide the four-dimensional description of the unpolarized cross sections.

\subsubsection{Projections}

\begin{table}[t]
\centering
\begin{tabular}{|c|c|c|c|c|}
\hline
Ion & 11+60 GeV & 3+20 GeV & 11+100 GeV & Polarization  \\\hline
p &  $9.3\times10^{40} \ {\rm cm}^{-2} $ &  $3.1\times10^{40}\ {\rm cm}^{-2}$  &
$3.1\times10^{40}\ {\rm cm}^{-2}$ & 1 \\\hline
D &$1.9\times10^{41}\ {\rm cm}^{-2} $  & $6.2\times10^{40}\ {\rm cm}^{-2}$ & $6.2\times10^{40}\ {\rm cm}^{-2}$ & 88\% \\\hline
$^3$He &$1.9\times10^{41}\ {\rm cm}^{-2}$  &$6.2\times10^{40}\ {\rm cm}^{-2}$  &$6.2\times10^{40}\ {\rm cm}^{-2}$  & 87.5\% \\\hline
\end{tabular}
\caption{Integrated luminosities, and the effective
polarization of the proton (neutron for D and $^3$He) in the
projections for different ion beams and EIC energy configurations.}
\label{tab:runtime}
\end{table}

In this section, we present the projected results of TSSA at an EIC.
Table~\ref{tab:runtime} summarizes the used run time distribution, luminosities, and effective
polarizations for different ion beams and energy configurations. 
In addition,  we assume polarizations of ion beams to be 
70$\%$ and an overall detecting efficiency of 50$\%$. 
The simulated data are binned according to different statistical precision for the TSSA 
measurement in different $Q^2$ regions. Different precisions are also chosen for 
different configurations. In particular, for both the 11+60 GeV and 11+100
GeV configurations, the statistical precision for each kinematic bin is set to be
about $2.0 \times10^{-3}$ for $Q^2<10$ GeV$^2$, and $4.0\times10^{-3}$ for $Q^2>10$ GeV$^2$.
For the 3+20 GeV configuration, $4.0\times10^{-3}$, and $5.0\times10^{-3}$ are chosen 
for $Q^2<10$ GeV$^2$ and $Q^2>10$ GeV$^2$, respectively. 

\begin{figure}[tbp]
\centering
\epsfig{file=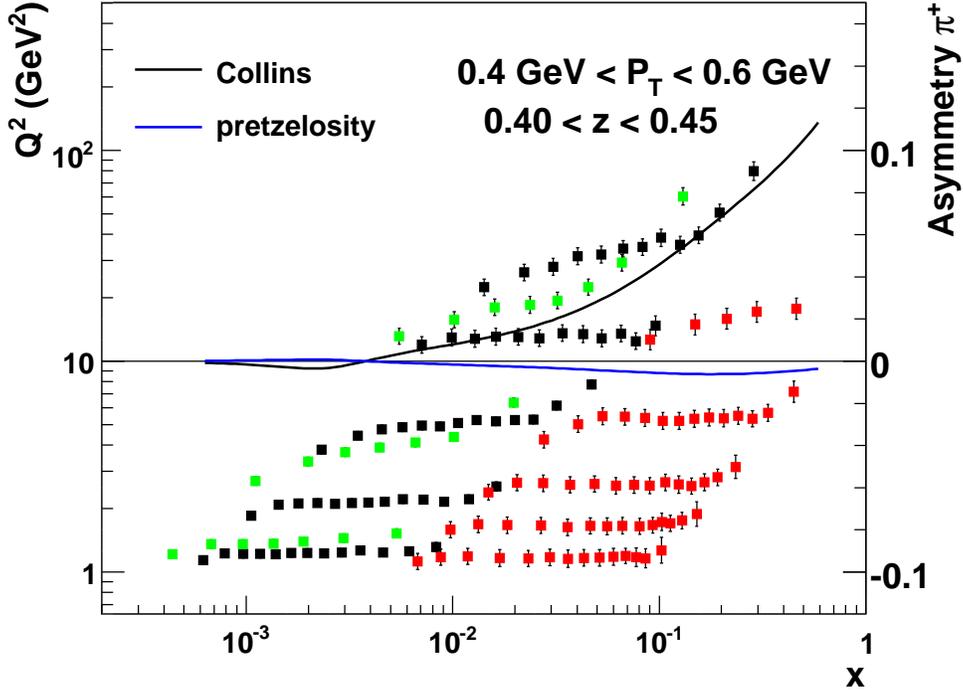,height=4.in,width=5.in}
\caption{Collins/pretzelosity asymmetry projection with proton on $\pi^+$ in a particular P$_T$ and 
  $z$ bin along with the calculated of asymmetries.
  The position of the dots are according to the $Q^2$ axis on the left and the $x$ axis,
  while the error bar of each dot is according to the scale of the
  asymmetry axis on the right.
  The calculated asymmetries are also according to the asymmetry axis. The black, green, and red dots 
  represent the 11+60 GeV, 11+100 GeV, and 3+20 GeV EIC configuration.}
\label{fig:proton_special_collins}
\end{figure}

\begin{figure}[tbp]
\centering
\epsfig{file=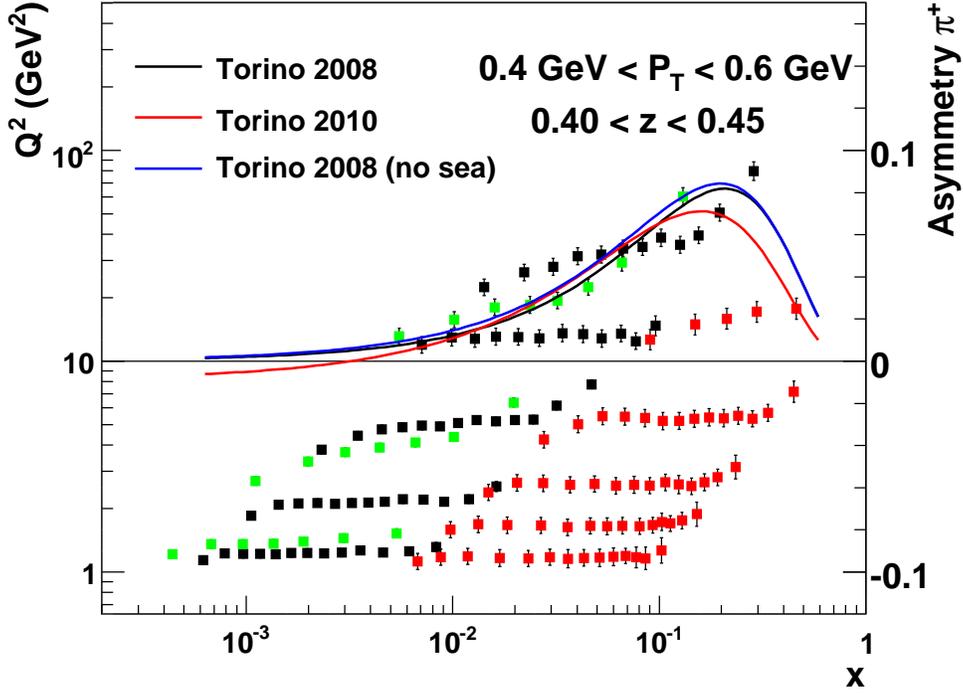,height=4.in,width=5.in}
\caption{Sivers asymmetry projection with proton on $\pi^+$ in a particular P$_T$ and 
  $z$ bin along with the calculated of asymmetries. The rest of the caption is the same as those in Fig.~\ref{fig:proton_special_collins}.}
\label{fig:proton_special_sivers}
\end{figure}

Fig.~\ref{fig:proton_special_collins} (Fig.~\ref{fig:proton_special_sivers}) shows the expected projection of $\pi^+$ Collins/pretzelosity (Sivers) asymmetry with a proton beam
at a high luminosity EIC in the kinematic bin of $0.4<z<0.45$ and $0.4~ {\rm GeV}<P_T<0.6~{\rm GeV}$. 
The $x$-axis 
represents Bjorken $x$, and the left $y$-axis is $Q^2$. The position of each point in the plot 
represents the position of the kinematic bin in the $x$-$Q^2$ phase space. The right $y$-axis 
is the asymmetry. The error bar of each point follows the right axis. Together with the projection, 
several asymmetry calculations are also presented. The codes to calculate the Collins and pretzelosity asymmetries are from~\cite{Huang:2007qn,She:2009jq}, and the Sivers asymmetry calculation is from ~\cite{Anselmino:2008sga}, and~\cite{Anselmino:2010sga} (red line).  
In the calculation, the PDF is from MRST2004 parametrization~\cite{Martin:2004ir},
and the FF is from Ref.~\cite{Kretzer:2001pz}. Ref.~\cite{She:2009jq} provides the Collins
and Pretzelosity distributions, in which the P$_T$ dependence is from Ref.~\cite{Anselmino:2008sj}. 
The Sivers TMD is according to Ref.~\cite{Anselmino:2008sga} and the recent result of Anselmino {\it et al.}, and the Collins FF is
according to Ref.~\cite{Anselmino:2008sj}. The calculated asymmetries also 
follow the right $y$-axis of the plot. 

\begin{sidewaysfigure}[tbp]
\centering
\epsfig{file=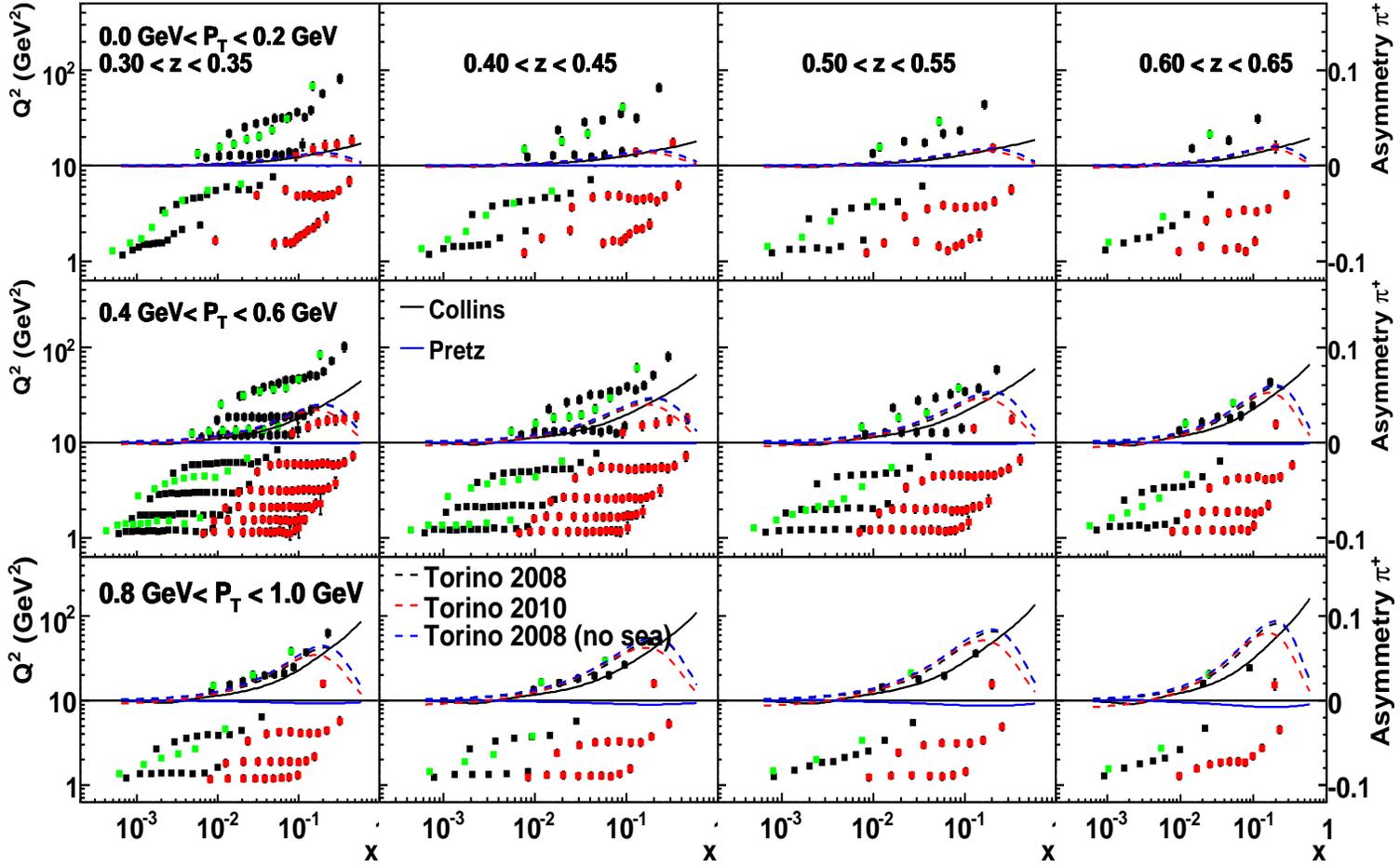,height=5.in,width=8.in}
\caption{4-D projections with proton on $\pi^+$ ($0.3<z<0.7$, 0 GeV/c$<P_T<$1 GeV/c).
The black, green, and red dots represent the 11+60 GeV, 11+100 GeV, and  3+20 GeV EIC configuration.
The rest of the caption is the same as Fig.~\ref{fig:proton_special_collins} and 
Fig.~\ref{fig:proton_special_sivers}.}
\label{fig:proton_pip}
\end{sidewaysfigure}

\begin{sidewaysfigure}[tbp]
\centering
\epsfig{file=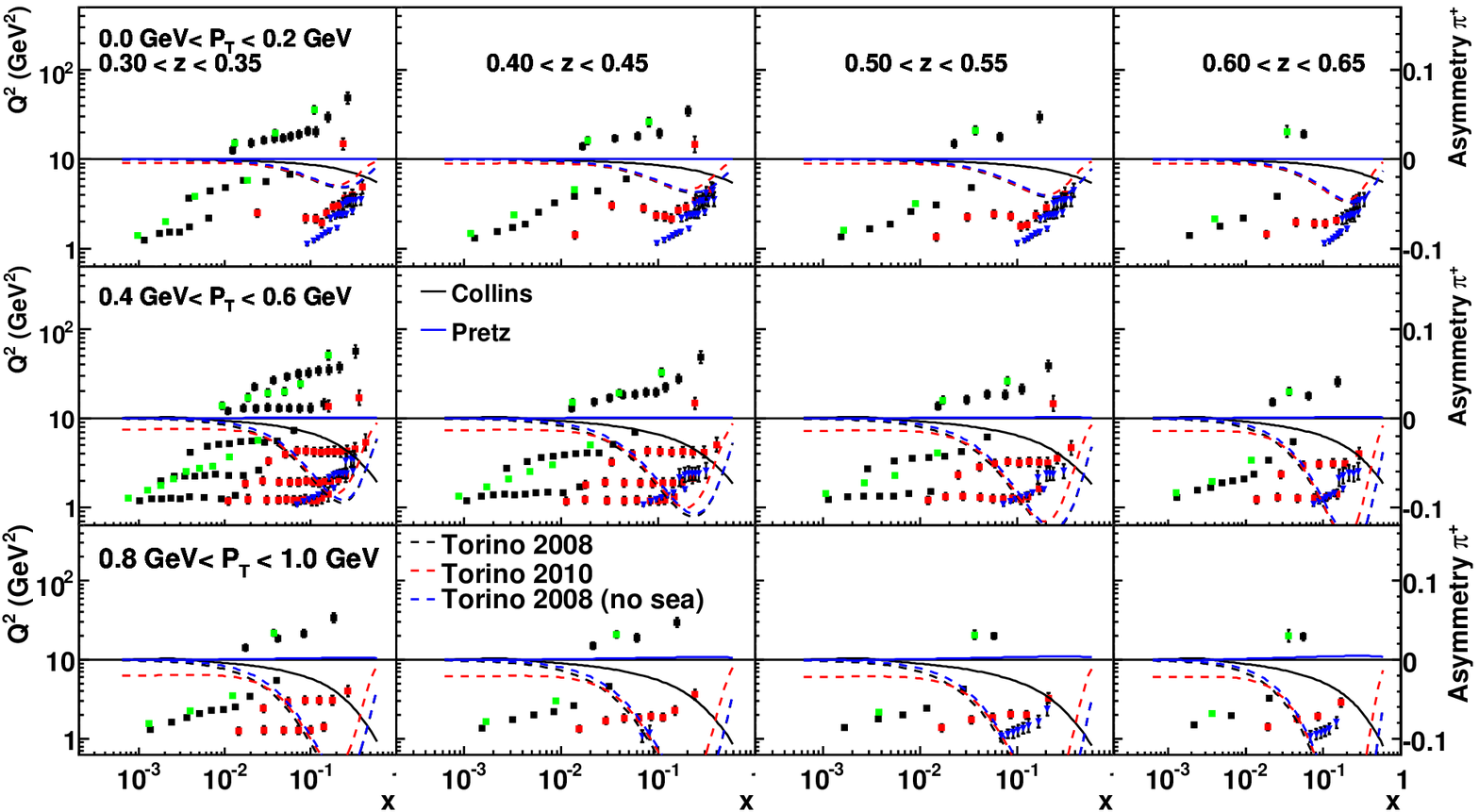,height=5.in,width=8.in}
\caption{4-D projection with $^3$He on $\pi^+$ ($0.3<z<0.7$, 0 GeV/c$<P_T<1$ GeV/c).
The blue dots represent the projection from the 11-GeV SoLID SIDIS experiment, which 
overlap with the red dots (3+20 GeV). The rest of the caption is the same as 
Fig.~\ref{fig:proton_special_collins} and Fig.~\ref{fig:proton_special_sivers}.}
\label{fig:3he_pip}
\end{sidewaysfigure}

\begin{sidewaysfigure}[tbp]
\centering
\epsfig{file=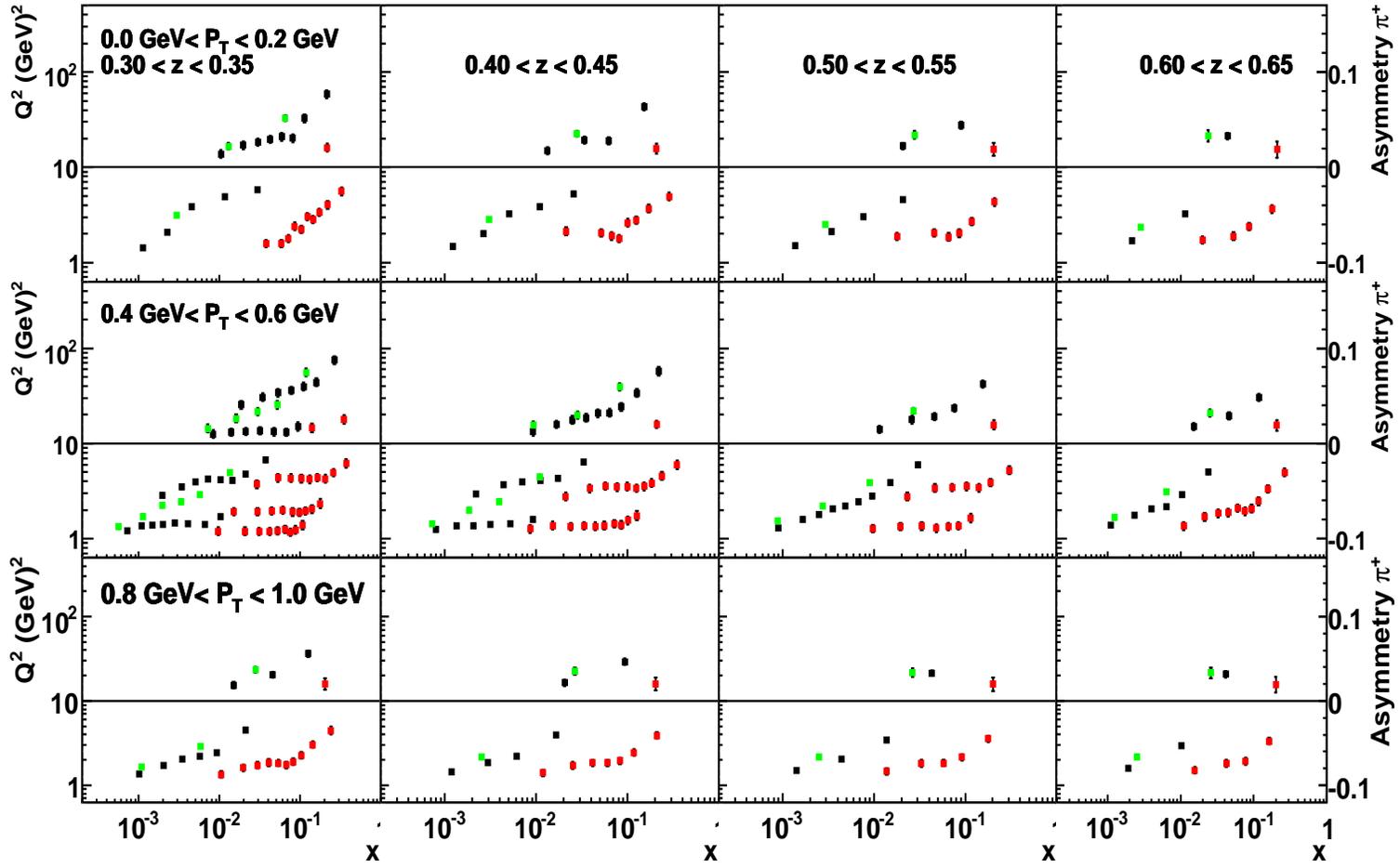,height=5.in,width=8.in}
\caption{4-D projection with proton on $K^+$ ($0.3<z<0.7$, 0 GeV/c$<P_T<$1 GeV/c). The rest of the caption 
is the same as Fig.~\ref{fig:proton_special_collins}.}
\label{fig:proton_kaon}
\end{sidewaysfigure}

The selected 4-D projections for the average of the Sivers, Collins and pretzelosity 
asymmetries for the entire phase space 
are shown in Fig.~\ref{fig:proton_pip} for 
$\pi^+$ with a proton beam. The entire $z$ coverage from 0.3-0.7 is divided into
8 bins (four $z$ bins are shown in Fig.~\ref{fig:proton_pip})
). We limit the projection at low $P_T$ region ($P_T<1$ GeV/c), where the $P_T$ coverage from 0 
to 1 GeV/c is divided into 5 bins (three $P_T$ bins are shown in Fig.~\ref{fig:proton_pip}). 
In Fig.~\ref{fig:proton_pip}, the central value of $z$ bins increases
from the left to the right. The central value of $P_T$ bins increases from the top to the bottom. 
In addition to the proton results, the neutron results can be obtained with polarized $^3$He and D beam.
The selected 4-D projections of the corresponding neutron results on $\pi^+$  using a polarized 
$^3$He beam are shown in 
Fig.~\ref{fig:3he_pip}~\footnote{There is a similar plot for Deuteron, which is not shown here.}. 
Together, the projected results for the 11 GeV SoLID SIDIS 
experiment~\cite{12GeV_transversity} are shown as blue points. The low energy EIC configuration (3+20 GeV)
would provide the data connecting the phase space from both the  fixed target experiment at 12 GeV JLab
and the high energy 11+60 GeV EIC configuration. Furthermore, with additional kaon particle identification, 
the kaon SIDIS results can provide additional handle for the flavor separation, since kaon results
would also tag the strange quark contribution from the sea. Fig.~\ref{fig:proton_kaon} shows the projected 
results of $K^+$ on proton in the selected 4-D phase space. Since the kaon rates are normally about one order
of magnitude lower than those of pions. The total number of points is significantly reduced. 

\begin{sidewaysfigure}[tbp]
\centering
\epsfig{file=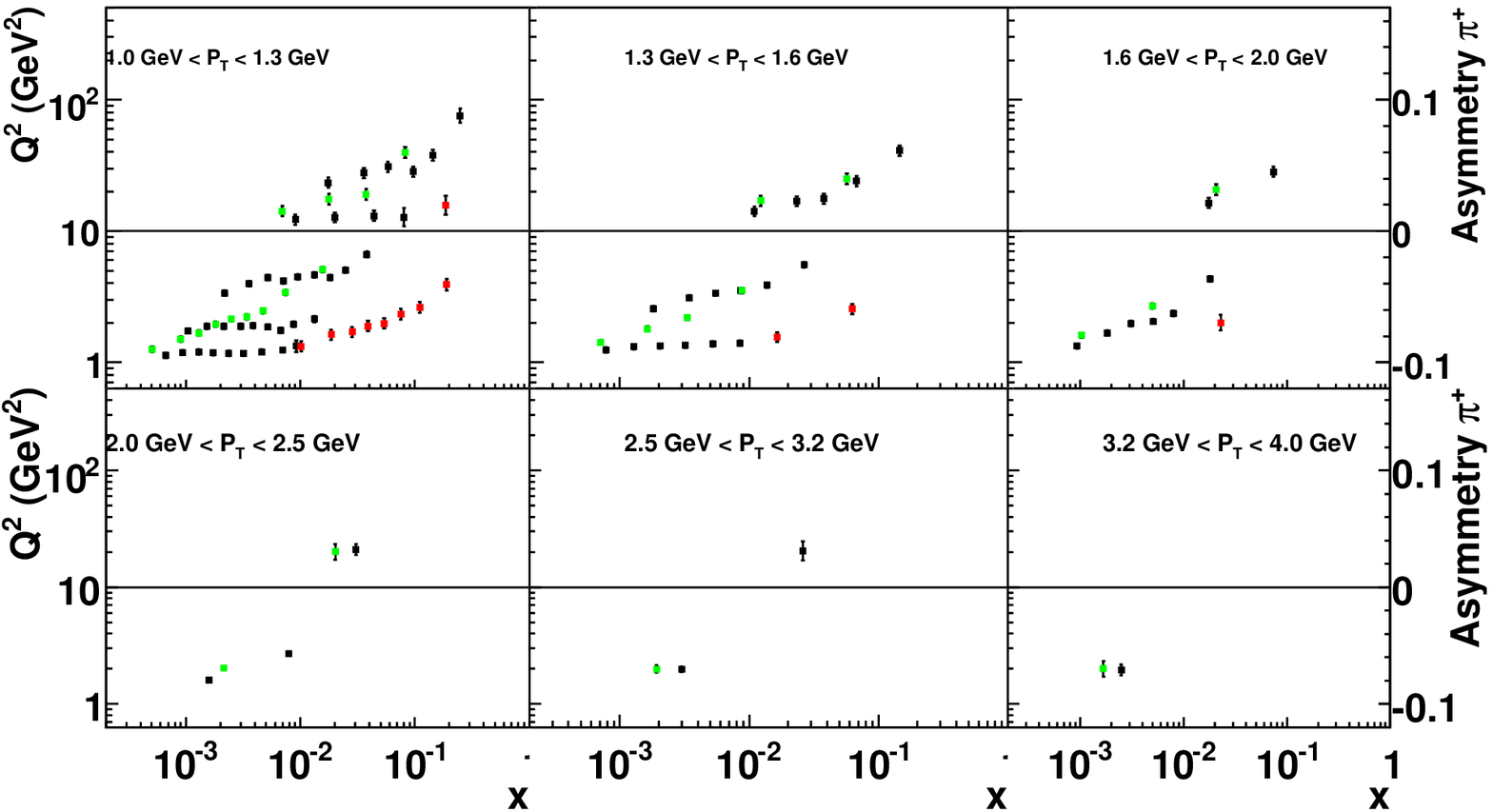,height=5.in,width=8.in}
\caption{4-D projection with proton on $\pi^+$ in one $z$ bin to show the
  P$_T$ dependence ($0.45<z<0.50$) in a range of 1 GeV/c$<P_T<$10 GeV/c (0 $<\log_{10}(P_T)<$ 1).}
\label{fig:pt_dep}
\end{sidewaysfigure}

In addition, the high center-of-mass energy $s$ at EIC would  enable the studies of TSSA in high 
$P_T$ region, where the twist-3 contribution will be large, and the intermediate $P_T$ region, where one
expects both the TMD and twist-3 formalism to work. 
In addition, a large $P_T$ coverage of TSSA would provide the chance of forming $P_T$ weighted 
asymmetry which is free of the Gaussian assumption of the transverse momentum dependence
for both the TMDs and FFs. Fig.~\ref{fig:pt_dep} shows, as an example, the P$_T$ dependence of the 4-D
projection with proton on $\pi^+$ in one $z$ bin. The number of points is limited at high $P_T$,
where the differential cross section decreases. 

From these projections, it is clear that the TSSA can be precisely mapped in the full 
$x$, $Q^2$, $z$ and $P_T$ 4-D phase space with a high luminosity EIC - 
a complete experiment (Table I) with a luminosity in excess of 10$^{34} cm^{-2} \cdot s^{-1}$ will need ~600 days of data taking. 
In particular, the EIC 
would facilitate the exploration of high Q$^2$ high $x$, and low $Q^2$ low $x$ phase space. 
Furthermore, the large coverage of $P_T$ would explore the TSSA in the high $P_T$ region
for the first time with SIDIS. The high luminosity is essential to realize the 
multi-dimensional mapping and extend the TSSA measurements to the unexplored regions (high $P_T$, 
high $Q^2$ etc.). 

\subsubsection{Projections for $P_T$-dependence of spin-azimuthal asymmetries}

\begin{figure}[tbp]
\begin{center}
\vspace{0.0cm}
\includegraphics[width=0.48\textwidth]{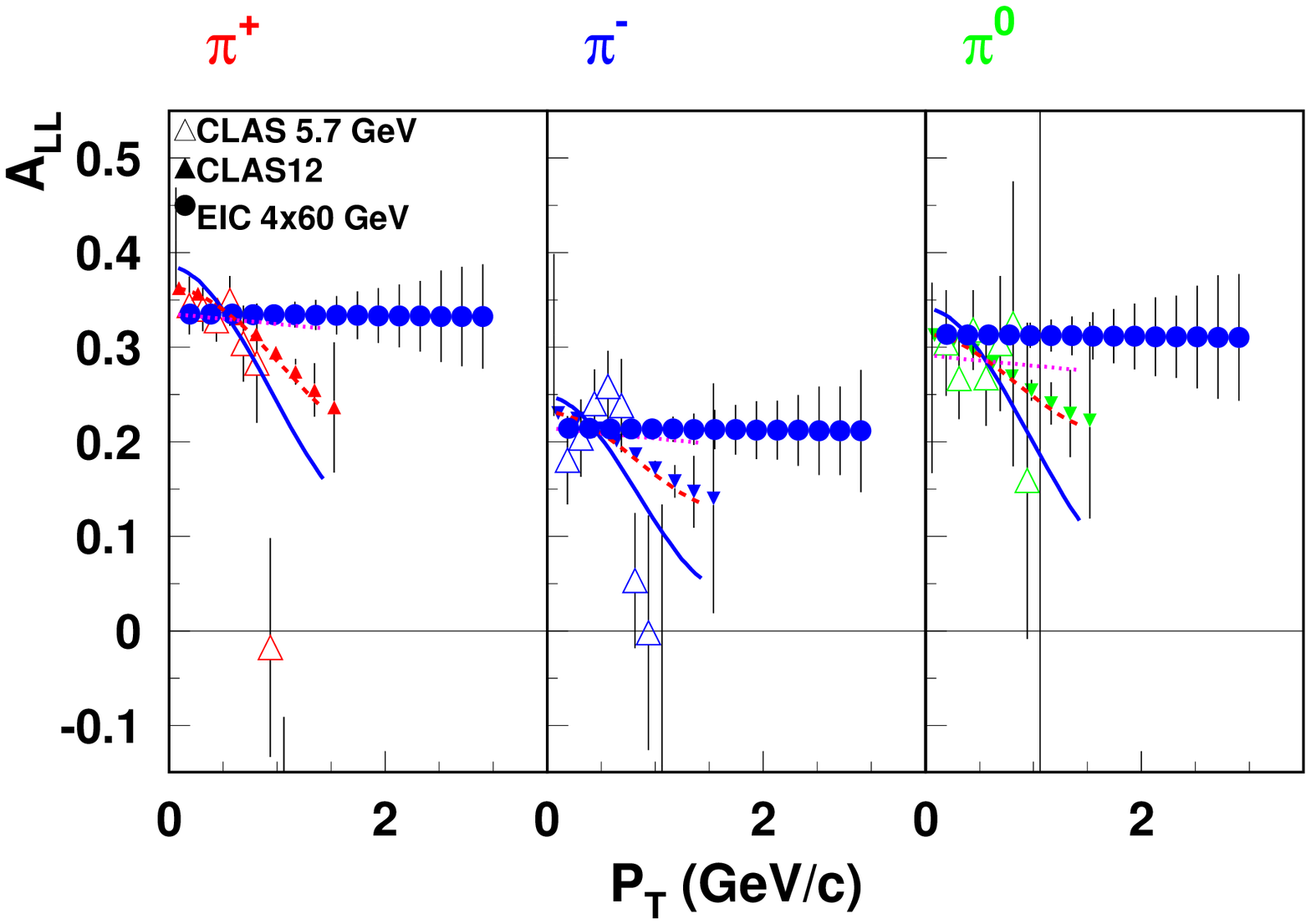}  \hfill 
\vspace{0.0cm}
\includegraphics[width=0.48\textwidth]{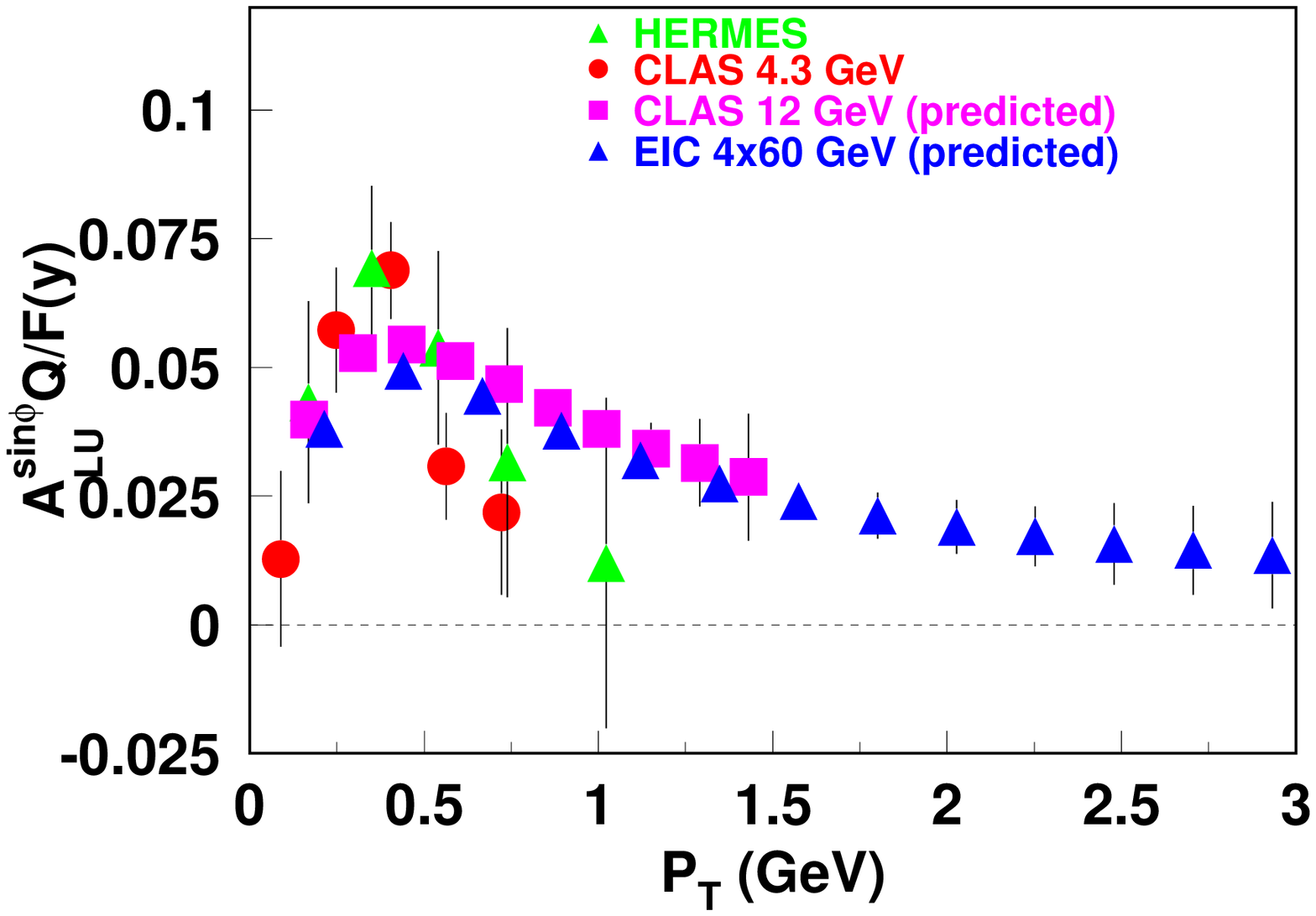}   
\end{center}
\caption{\small Double-spin asymmetry, $A_{LL}$, for positive pion
production, using 4 GeV electrons and 60 GeV protons (100 days at 10$^{34}$ cm$^{-2}$ sec$^{-1}$),  as a
function of  $P_T$ compared to published data from CLAS
\cite{Avakian:2010ae} and projected CLAS12 measurements \cite{PACall} (left).
The right plot shows projections for the same running conditions for the
higher-twist lepton spin
asymmetry compared to published data
from CLAS \cite{Avakian:2003pk} and
HERMES~ \cite{Airapetian:2006rx} and projected CLAS12 \cite{PACauu} in one
$x,z$ bin ($0.2<x<0.3$, $0.5<z<0.55$).}
\label{Allpt_dep}
\end{figure}

Significantly higher, compared to JLab12, $P_T$ range accessible at EIC would allow for studies
of transverse momentum dependence of different distribution and fragmentation functions as well as 
transition from TMD regime to perturbative regime.
Measurements of double-spin asymmetries as a function of the final hadron transverse momentum at EIC will
extend (see Fig. \ref{Allpt_dep}) measurements at JLAB12~\cite{PACauu,PACall}
 to significantly higher $P_T$ and $Q^2$ 
allowing comparison with calculations  performed in the perturbative limit~\cite{Zhou:2009jm}.
Extending measurements of $P_T$-dependent observables to significantly
lower $x$ will provide access to transverse momentum dependence of quarks beyond the valence region.
Much higher $Q^2$ range accessible at EIC would allow for
studies of $Q^2$-dependence
of different higher twist SSAs, which, apart from providing important
information on quark-gluon
correlations are needed for understanding of possible corrections from
higher twists to leading
twist observables.

\subsection{Transverse Single Spin Asymmetry for $D$ ($\bar{D}$) Mesons}

As discussed in Sec. I, the transverse single spin asymmetry of 
semi-inclusive neutral $D$ meson production at large transverse momenta 
$P_T$ would open a new window to the unexplored 
tri-gluon correlation functions. In this section, we discuss the potential 
of an EIC in this measurement using PYTHIA simulation~\cite{pythia}, which has been
compared with PEPSI in the inclusive electron scattering, and with PEPSI and 
our newly developed cross section weighting approach for the SIDIS 
pion production in Sec. C. Here, we compare our projections with calculations based on an earlier paper on this subject~\cite{Kang:2008qh}. New calculations 
based on the latest development~\cite{Beppu:2010qn} is ongoing.
The mass of the charm quark is taken as 1.65 GeV 
and the rest of the input parameters are from Ref.~\cite{ref:elke_pc}.
The main channels of interest are
\begin{eqnarray}
D^0(c\bar{u}) &\rightarrow& \pi^+ (u\bar{d})K^-(s\bar{u}) \\
\bar{D^0}(\bar{c}u) &\rightarrow& \pi^-(\bar{u}d) K^+ (u \bar{s})
\end{eqnarray}
where the branching ratio is 3.8 $\pm$ 0.1 \%~\cite{Nakamura:2010zzi}. The
two other main decay channels:
$\bar{D}->K^+\pi^-\pi^0$ and
$\bar{D}->K^+\pi^-\pi^-\pi^+ $
are under investigation.

\begin{figure}[tbp]
\centering
\epsfig{file=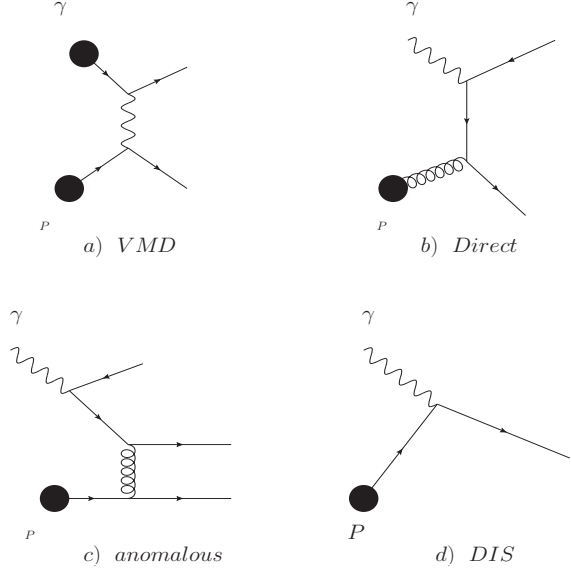,height=3.in,width=3.in}
\caption{Four diagrams of hard $\gamma p$ interactions modeled in PYTHIA.}
\label{fig:four}
\end{figure}

\begin{figure}[tbp]
\centering
\epsfig{file=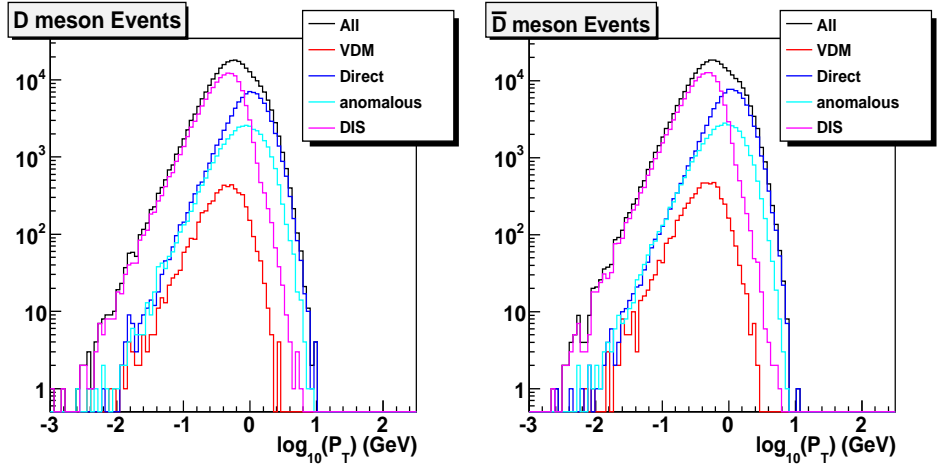,height=2.5in,width=5.in}
\caption{Contributions from four $\gamma p$ interaction processes modeled in 
PYTHIA vs. $P_T$ for $D$ ($\bar{D}$)-meson production.
At large $P_T$, the ``direct'' production process dominates. 
The largest contamination 
is from the ``anomalous'' production at $P_T > 1$ GeV.}
\label{fig:cont1}
\end{figure}

\begin{figure}[tbp]
\centering
\epsfig{file=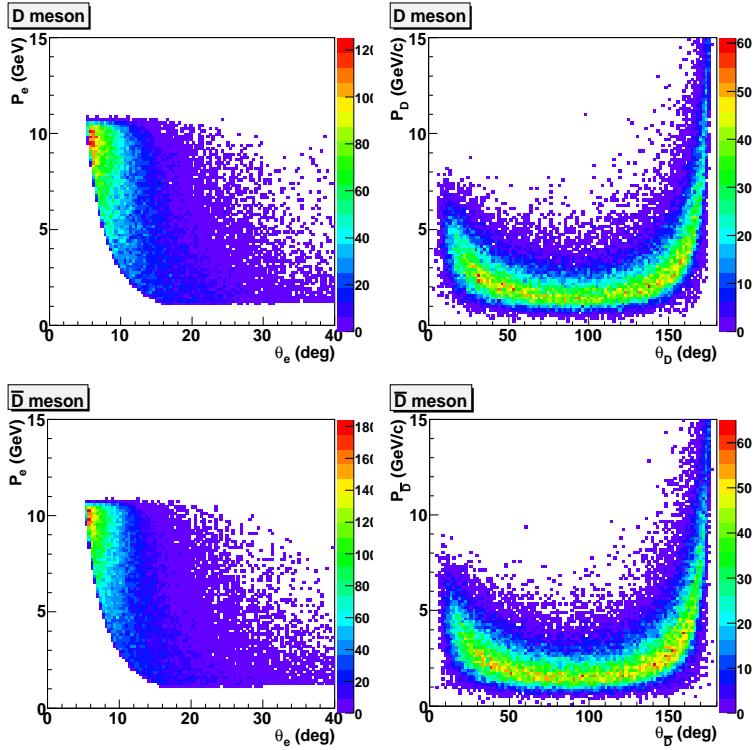,height=4.in,width=4.in}
\caption{Top (bottom) two panels show the momentum vs. polar angle distribution for electron and $D$
  ($\bar{D}$) in the lab frame.}
\label{fig:phase1}
\end{figure}

\begin{figure}[tbp]
\centering
\epsfig{file=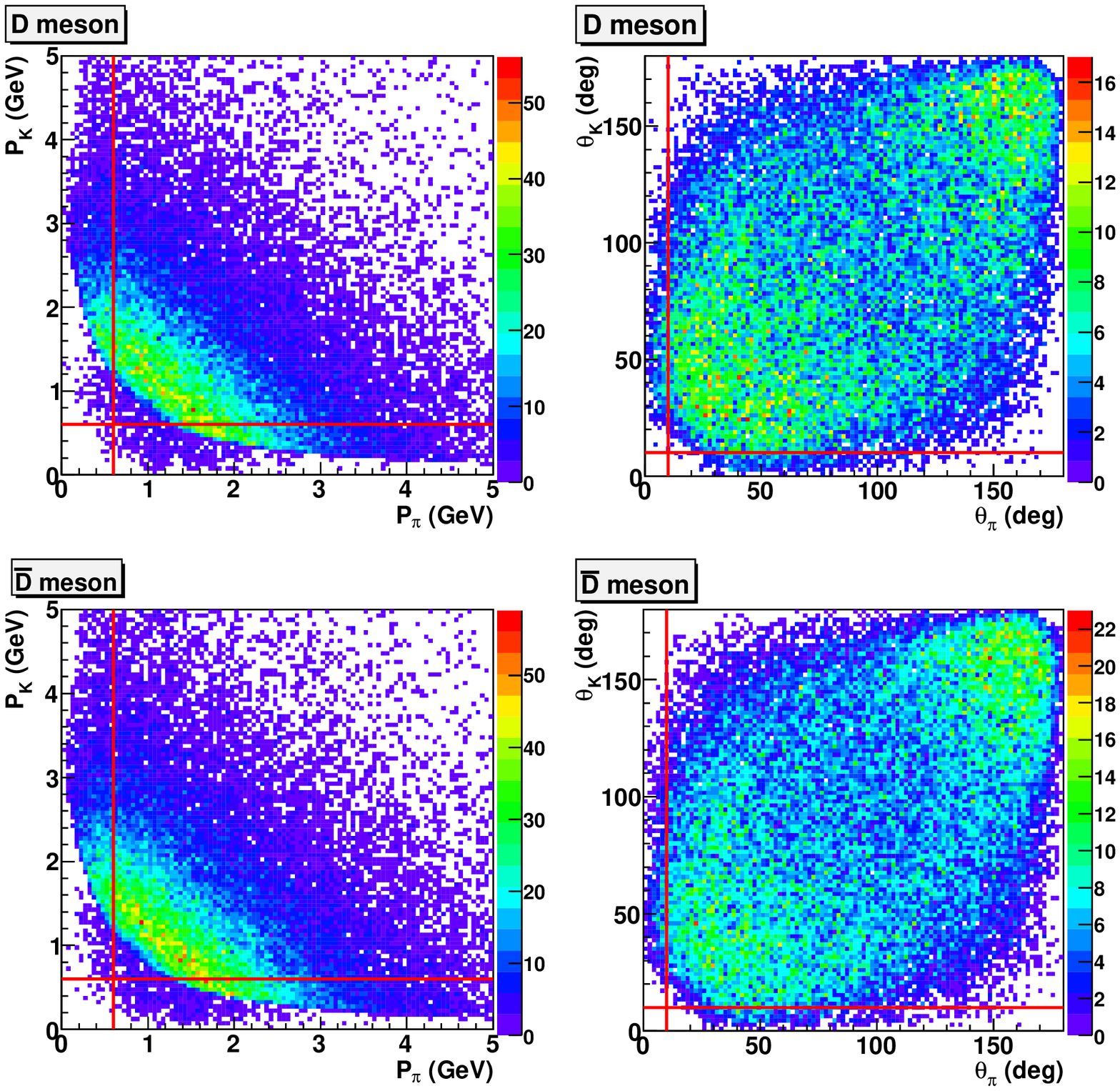,height=4.in,width=4.in}
\caption{Left two panels show the momentum of $\pi$ vs. that of $K$ from $D$ ($\bar{D}$) meson
decay in the lab frame. Right two panels show the polar angle of $\pi$ vs. that of $K$. Red lines 
show the expected momentum and polar angle cut-off due to a realistic 
acceptance 
of an EIC detector.}
\label{fig:phase2}
\end{figure}

The simulated phase space is limited in $0.05 < y < 0.9$ and 
$Q^2 > 1.0$ GeV$^2$ for the 11+60 EIC configuration with a proton beam. 
The main channel is the ``direct'' production channel, which is one of 
the four reaction mechanisms for producing neutral $D$ mesons (vector meson dominance (VMD), anomalous, direct and DIS) as shown 
in Fig.~\ref{fig:four} modeled in PYTHIA~\cite{Schuler:1992dt} for 
the hard $\gamma-p$ interactions. The $P_T$ distributions of simulated $D$ and $\bar{D}$ events are shown in Fig.~\ref{fig:cont1} for all four mechanisms. 
The ``direct'' process dominates the production at $P_T > 1$ GeV, and
the largest contamination is from the ``anomalous'' process. In the following 
studies, additional cuts of $z > 0.15$ and $P_T > 1$ GeV are applied. 
Fig.~\ref{fig:phase1} shows the momentum-polar angle distribution of electron
and $D$ meson in the lab frame.  Fig.~\ref{fig:phase2} shows the momentum-polar angle distribution of the $\pi$ and $K$ from the $D$ ($\bar{D}$) meson decay. 
The minimum momentum cut-off is chosen to be 0.6 GeV, and the minimum polar 
angle is chosen to be 10 degrees.

\begin{figure}[tbp]
\centering
\epsfig{file=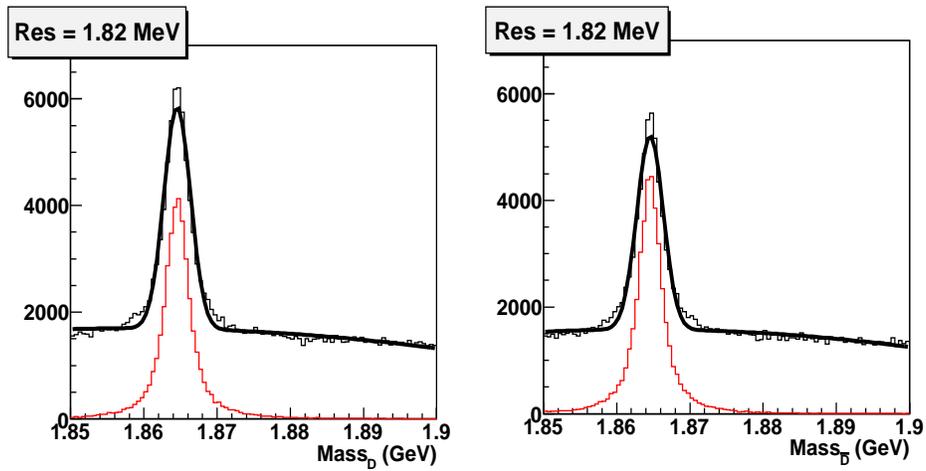,height=2.5in,width=5.in}
\caption{The reconstructed invariant mass spectra of $D$ and $\bar{D}$ are shown in the left and right panel,
respectively. The resulting resolution on invariant mass is about 1.8 MeV. }
\label{fig:inv}
\end{figure}

In order to make projections, the $D$ ($\bar{D}$) mesons have to 
be reconstructed from the measured $\pi$ and $K$ for each of the generated events. The signal-to-background ratio would 
strongly depend on the detector resolution. In this study, 
the momentum resolution is assumed to be 
$0.8\% \cdot \frac{p}{10 GeV}$~\cite{ref:ent_pc}. The resolutions of polar and azimuthal angle 
are assumed to be 0.3 and 1 mr, respectively. The resulting resolution of the reconstructed invariant mass of the $D$ meson 
is 1.8 MeV as shown in Fig.~\ref{fig:inv}. In this case, the overall signal-to-background 
ratio is about 1.6 to 1. The background under the D meson invariant mass peak is due to the random 
coincidence of unrelated $\pi$ 
and $K$ in the final states after applying all cuts. The dilution of the background to the measured 
asymmetry is assumed to be:
\begin{equation}
\delta A = \frac{1}{\sqrt{S}} \cdot \sqrt{\frac{S+B}{S}}
\end{equation}
where $S$ and $B$ represent the signal and background within the invariant mass cut 
($1.86< M_D < 1.87$ GeV). The dilution due to the non-''direct'' D meson production is applied in the same manner as in the final projection. 

\begin{figure}[tbp]
\centering
\epsfig{file=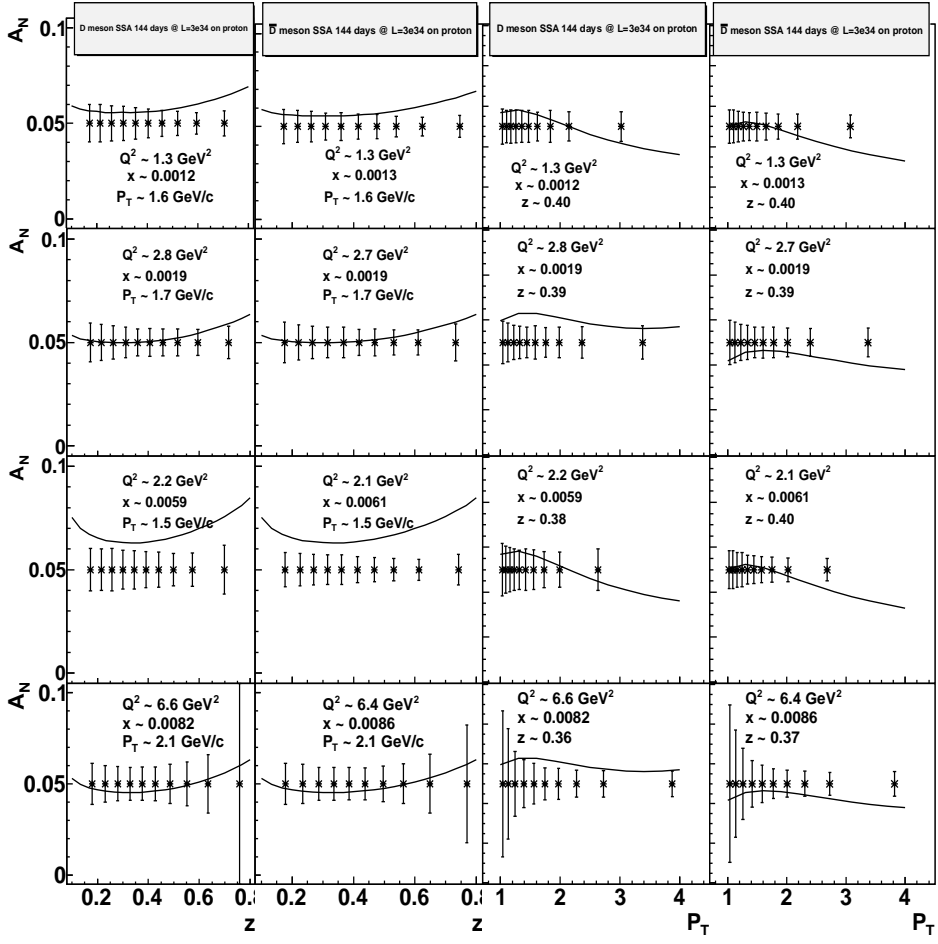,height=5.in,width=5.in}
\caption{The projected results on transverse SSA of $D$ ($\bar{D}$) meson production. 
The data are binned 2-by-2 in terms of $x$ and $Q^2$. Within each $x$-$Q^2$ bin, 
the projections are either plotted with $z$ or $P_T$. The central kinematics are also listed in each 
plot. The first (second) column shows the projected results vs. $z$ for $D$ ($\bar{D}$) meson. 
The third (fourth) column shows the projected results vs. $P_T$ for $D$ ($\bar{D}$) meson.}
\label{fig:proj}
\end{figure}

Fig.~\ref{fig:proj} shows the projection for $D$ ($\bar{D}$) transverse SSA measurement for a running time of 
 144 days of proton beam at a luminosity of $3\times 10^{34}/cm^2 \cdot s^{-1}$. 
In this plot, the effect of $\pi$ and $K$ decay is included. 
The overall detection efficiency for this triple
coincidence process is assumed to be 60\%. The polarization of the proton beam is assumed to be 80\%.
An additional factor of $\sqrt{2}$~\footnote{Here, we adopt the first order approximation. As illustrated in previous section, such a factor, which is close to $\sqrt{2}$,  would depend on the azimuthal angular coverage and event distribution.}
 is included in order to take into account the loss of precision in the angular
separation. The data are binned 2-by-2 in terms of $x$ and $Q^2$. Within each $x$-$Q^2$ bin, 
the projections are either plotted with $z$ or $P_T$. The central kinematics are also listed in each 
of the panel. We show also theoretical predictions from Ref.~\cite{Kang:2008qh,Kang:2008ey,ref:kang_pc}.

\subsection{Requirements on an EIC detector}

In this section, we summarize the requirements on an EIC detector from SIDIS processes.
\begin{itemize}
\item {\bf Electron detection:} \\
  As shown in Fig.~\ref{fig:elepth}, with the DIS cut of $Q^2 > 1$ GeV$^2$, 
there is no need to 
  cover the extreme forward angle. The minimum polar angle coverage can be estimated by:
  \begin{equation}
    \theta_{min} \sim 2 \arcsin(1/P_e^0),
  \end{equation}
  where $P_e^0$ is the momentum of the incident electron beam.
  In addition, most of the scattered electrons have large momenta. Therefore, 
  it is desirable to have:
  \begin{equation}
    \frac{\delta p}{p } < 1\% \cdot \frac{p}{10 {\rm GeV}}
  \end{equation}
  in order to achieve a good resolution of Bjorken $x$~\cite{ref:ent_pc}.
\item {\bf SIDIS $\pi$ or $K$ production at $P_T<1$ GeV:} \\
  As shown in Fig.~\ref{fig:hadpth}, leading hadrons span 
  a large polar angular coverage in the lab frame. The momenta of
  most leading hadrons would be limited between 0.8 GeV and 7 GeV. 
  Therefore, the separation of $p/\pi/K$ between 2.5 and 170 degrees and 
  for momenta smaller than 7 GeV is essential to the success 
  of the SIDIS program. In addition, a lower momentum cut-off for hadrons
  will enhance the overall acceptance. 
\item {\bf SIDIS $\pi$ or $K$ production at $P_T>1$ GeV:} \\
  As shown in Fig.~\ref{fig:hpt_pth_had}, the high $P_T$ events favor a large 
  hadron momentum in the lab frame. Therefore, the separation of $p/\pi/K$ for momenta
  larger than 7 GeV would be very useful for the high $P_T$ SIDIS physics. 
\item {\bf SIDIS $D$ or $\bar{D}$ production at $P_T>1$ GeV:}\\
  As shown in Fig.~\ref{fig:phase2}, the momentum of the $\pi$ and $K$ (decay products
  of $D$ meson) is in general smaller than 5 GeV. Therefore, a separation of $p/\pi/K$ 
  between 2.5 and 180 degrees and for momenta smaller than 5 GeV is 
  adequate for the identification of $D$ or $\bar{D}$ meson for the transverse SSA
  physics. The more challenging requirement is on the detector resolution.
  As shown in Fig.~\ref{fig:inv}, a $0.8\% \cdot \frac{p}{10 GeV}$ momentum resolution, a 
  0.3 mrad polar and a 1 mrad azimuthal angular resolutions lead to a 1.8 MeV resolution
  of the reconstructed invariant mass of the $D$ meson. A better detector resolution will lead to 
  a better signal-to-noise ratio, which then leads to a better precision.
    
\item {\bf Luminosity and Energy Coverage:} \\
  Due to the multi-dimensional nature of SIDIS processes, the 4-D ($x$, $Q^2$, $P_T$ and $z$) 
  mapping of transverse SSA is essential for the success of TMD physics through SIDIS at an EIC. 
  In addition, from Fig.~\ref{fig:energies}, it is essential to cover a few different 
  C.M. energies to reach ultimate mapping of asymmetries in $x$-$Q^2$ phase space. 
  Therefore, it is essential to have $L>1\times 10^{34}/cm^2 \cdot s^{-1}$ at multiple $s$ values (e.g. 11+100, 
  11+60, and 3+20 configurations).
\end{itemize}

In this paper, we have summarized 
a recent workshop held at Duke University 
on Partonic Transverse Momentum in Hadrons: 
Quark Spin-Orbit Correlations and Quark-Gluon Interactions.
The workshop participants identified the SSA measurements in SIDIS as a golden program to study TMDs in both the sea and valence quark regions as well as to study the role of gluons, with the 
Sivers asymmetry measurements as examples.
A high-intensity EIC with a wide center-of-mass energy range will allow for studies of TMDs in multi-dimensional phase space with high precisions in both the valence quark region at high $Q^2$ and the unexplored sea quark region, and allow for studies of tri-gluon correlation functions.
Such studies will greatly advance our knowledge about the structure of the nucleon in three dimensions and transverse spin physics.

\section{acknowledgment}
This work is supported in part by 
the U.S. Department of Energy under contracts, DE-AC05-84ER40150,
modification No. M175, under which the Southeastern Universities
Research Association operates the Thomas Jefferson National Accelerator
Facility, and DE-FG02-03ER41231 (H.G.). We also thank Jefferson Science 
Associates (JSA), Jefferson Lab, and the Triangle Universities Nuclear 
Laboratory for the support of this workshop.

\bibliography{duke}

\end{document}